\documentclass[lettersize,journal,twocolumn]{article}
\usepackage{cite}
\usepackage{amsmath,amssymb,amsfonts}
\usepackage{algorithmic}
\usepackage{graphicx}
\usepackage{bm}
\usepackage{subfig}
\usepackage{caption}
\usepackage{balance}
\usepackage{mathtools}
\usepackage{booktabs}
\usepackage{float}
\usepackage{changepage}
\usepackage[left=1.60cm,right=1.60cm,top=1.60cm,bottom=2.40cm]{geometry}
\usepackage{rotating}
\graphicspath {{figures/}}
\usepackage{wrapfig}
\usepackage{placeins}
 \usepackage{hyperref}
\hypersetup{
    colorlinks=true,
    citecolor=blue     
    }
\usepackage{multicol, blindtext}
\usepackage[table]{xcolor}
\usepackage{pgfplots}
 \usepackage{wrapfig}
 \usepackage{xcolor}
 \usepackage{tabularx}
 \usepackage{multirow}
\usepackage{bigstrut}
\usepackage{authblk}
\usepackage[pdftex]{lscape}
\usepackage{rotating}
\usepackage{multicol}

\usepackage{bm,array}
\usepackage{color, colortbl}
\definecolor{Gray}{gray}{0.9}
\definecolor{Grayy}{gray}{0.7}
\usepackage{makecell}
\usepackage{nicematrix}
\usepackage{textcomp}
\providecommand{\keywords}[1]{\textbf{\textit{Keywords---}} #1}

\begin{document}
\title{Higher Order Dynamic Mode Decomposition: from Fluid Dynamics to Heart Disease Analysis}
\author[1]{N. Groun}
\author[2]{M. Villalba-Orero}
\author[2]{E. Lara-Pezzi}
\author[3]{E. Valero}
\author[3]{J. Garicano-Mena}
\author[3]{S. Le Clainche}
\affil[1]{ETSI Aeronáutica y del Espacio and ETSI Telecomunicación - Universidad Politécnica de Madrid, 28040 Madrid, Spain}
\affil[2]{Centro Nacional de Investigaciones Cardiovasculares (CNIC), C. de Melchor Fernández Almagro, 3, 28029 Madrid, Spain}
\affil[3]{ETSI Aeronáutica y del Espacio - Universidad Politécnica de Madrid, 28040 Madrid, Spain and Center for Computational Simulation (CCS), 28660 Boadilla del Monte, Spain}
\date{}                     
\setcounter{Maxaffil}{0}
\renewcommand\Affilfont{\itshape\small}

\maketitle

\begin{abstract}
In this work, we study in detail the performance of \textit{Higher Order Dynamic Mode Decomposition} (HODMD) technique when applied to echocardiography images. HODMD is a data-driven method generally used in fluid dynamics and in the analysis of complex non-linear dynamical systems modeling several complex industrial applications. In this paper we apply HODMD, for the first time to the authors knowledge, for patterns recognition in echocardiography, specifically, echocardiography data taken from several mice, either in healthy conditions or afflicted by different cardiac diseases. We exploit the HODMD advantageous properties in dynamics identification and noise cleaning to identify the relevant frequencies and coherent patterns for each one of the diseases. The echocardiography datasets consist of video loops taken with respect to a long axis view (LAX) and a short axis view (SAX), where each video loop covers at least three cardiac cycles, formed by (at most) 300 frames each (called snapshots). The proposed algorithm, using only a maximum quantity of 200  snapshots, was able to capture two branches of frequencies, representing the heart rate and respiratory rate. Additionally, the algorithm provided a number of modes, which represent the dominant features and patterns in the different echocardiography images, also related to the heart and the lung. Six datasets were analyzed: one echocardiography taken from a healthy subject and five different sets of echocardiography taken from subjects with either Diabetic Cardiomyopathy, Obesity, SFSR4 Hypertrophy, TAC Hypertrophy or Myocardial Infarction. The results show that HODMD is robust and a suitable tool to identify characteristic patterns able to classify the different pathologies studied.
\end{abstract}


\section{Introduction}
\label{sec:introduction}
Data analysis is a rapidly evolving field that allows us to discover useful information from raw data and support decision-making in different domains such as marketing~\cite{ringel2016visualizing},\cite{jacobs2016model}, cyber-security~\cite{foroughi2018data},\cite{sarker2020cybersecurity},\cite{thuraisingham2016data}, sports~\cite{leung2014sports},\cite{maszczyk2014application},\cite{mccullagh2010data}, health care\cite{ryan2007breast},\cite{zhang2017analytical},\cite{islam2018brain}  etc. In health-care in particular, medical imaging plays a crucial role, as it enables medical practitioners to identify diseases in its early stages, to provide accurate diagnoses and to plan and guide the optimal treatment for every situation. Medical imaging field is being profoundly affected by the technological revolution brought forward by increasingly sophisticated electronic devices and the continuous growth of computing power. As a consequence, medical imaging has become a data intensive field: optimized tools grounded in the data science discipline are necessary to reap the full potential of the wealth of data available. Data analysis methods, such as model decomposition or neural networks, are being approached more and more because of the fact that they are purely data-driven and they do not require any knowledge of the underlying equations, which make it easier for biologists, medical scientists and engineers to collaborate to obtain accurate results.\\ 
In different fields, large data matrices will be produced by complex systems\cite{ottino2003complex}. For example, an experiment will result in a matrix where the columns include the measurements obtained at different time instants; similarly, a number of images can be reshaped into vectors and arranged in  matrix form, where each column will represent a shot (snapshot) of a video. When it comes the medical field, medical imaging occupies a large percentage of medical data. Usually in medicine a large number of images need to be examined but only a few will be expected to show abnormalities, so data science or data analysis methods can be considered as an assistant to a radiologist to detect lesions and make smart decisions. Furthermore, the increasing in the numbers of sophisticated electronic devices, the new acquisition and the storage of medical images containing relevant information, reveals the need of finding new tools for data analysis capable to extract relevant information from these databases.\\
Generally, when it comes to  medical imaging, machine learning (ML) and Artificial intelligence (AI) have been highly used. In the health care industry, heart diseases are the number 1 cause of death globally. According to the American Heart Association, in 2017,  17.8 million deaths  were  attributed to cardiovascular diseases\cite{virani2020heart} and  it is considered the single largest cause of death in the world taking more than a third of all deaths according to The Global Burden Of Disease in 2004\cite{mathers2008global}. These numbers are putting a lot of pressure on health care systems all around the world. In parallel with the increase in the number of patients, the associated data is also growing, and it is very difficult for traditional methods to keep up with the fast growth of data. This begs for the intervention of data science techniques to help the medical field to overcome some of the obstacles they face. Many researchers have used ML and AI to help in the diagnosis of heart diseases. For example, Lelieveldt \textit{et al.}\cite{10.1007/3-540-45729-1_47} did an impressive work while trying to extend the 2D time Active Appearance Motion Model (AAMM), establishing a time-continuous segmentation of cardiac image sequences. They tested the approach on short-axis cardiac magnetic resonance imaging (MRI) with total of 1200 image frames from 25 subjects, 15 normal subjects and 10 myocardial infarction patients and four-chamber echocardiographic image sequences from 129 unselected patients. Although their method performed slightly more accurately for MRI than for echo-cardiograms it still generates a nearly perfect time-continuous segmentation results, which are consistent
with cardiac dynamics. Arsanjan \textit{et al.}\cite{worden2015second} used logistic regression to improve a combination of classifiers in order to diagnose obstructive  coronary artery disease (CAD) using single-photon emission computerized tomography  (SPECT) images. Berikol \textit{et al.}\cite{berikol2016diagnosis} also has shown that ML, in particular, support vector machines were extremely successful in predicting acute coronary syndrome for 228 patients.  Xulei Qin \textit{et al.}\cite{qin2013extracting} proposed an extraction technique to automatically detect the cardiac myofiber orientations from high frequency ultrasound images. This method was tested on both phantom and pig hearts and showed satisfying results in both cases (see also\cite{611346},\cite{WAITER199999},\cite{huang2011image},\cite{9082648}). \\ 
Furthermore, starting from a certain point in the mid 2000s, approaches based on matrix decomposition and data-driven methods began to gain recognition in medical analysis field. Veltri \textit{et al.}~\cite{VELTRI201057} used proper orthogonal decomposition (POD) to help the diagnosis of kidney diseases. When analyzing six datasets, four belong to patients affected by renal pathologies and two for healthy patients, POD was able to underline the regions of the organ interested by the troubles allowing to analyze them independently (see also~\cite{grinberg2009analyzing},~\cite{fathi2018denoising}). Meanwhile, Nika \textit{et al.}(\cite{NIKA2014527},~\cite{10.1117/1.JMI.1.2.024502}) merged principal component analysis (PCA) in their  algorithm named EigenBlockCD, to detect changes in serial MR images of the brain. The main idea of their algorithm was to perform a local image registration to identify important structural changes ignoring unimportant changes related to misalignment, noise and acquisition-related artifacts, supported with PCA as a feature extraction tool (by emphasizing most significant features within the images) and a dimensional reduction tool (to reduce the dimensionality of the dictionary and hence increase the computational efficiency). Also with a different approach, PCA combined with the least absolute shrinkage and selection operator (LASSO), Klyuzhin \textit{et al.}~\cite{klyuzhin2018data} came up with data-driven, voxel-based analysis of brain positron emission tomography (PET) images, where they worked on identifying voxel covariance patterns using PCA and then using LASSO to combine several patterns to construct models that predict clinical disease metrics from imaging data (see also~\cite{tirunagari2019functional},~\cite{fathi2020time},~\cite{balasubramanian2007automatic}). Moreover, dynamic mode decomposition (DMD), originally used as a tool to identify patterns  in fluid dynamics (see e.g. \cite{rowleyEtAlJFM2009},\cite{pof2019},\cite{schmid2010dynamic},\cite{energies2020}), was reused in the medical field. Tirunagari \textit{et al.}~\cite{tirunagari2017movement} incorporated different extensions of DMD to develop a novel automated, registration-free movement correction approach for kidney dynamic contrast-enhanced magnetic resonance imaging (DCE-MRI). Their method was tested on ten different datasets, while comparing the images produced by their approach with the original dataset, the results showed the elimination of 99\% of mean motion magnitude. Fua \textit{et al.}~\cite{fu2020novel} exploited DMD for the diagnose of Parkinson's disease. Specifically, they applied the DMD algorithm to extract spatio-temporal patterns of neurotransmitter changes due to neurodegeneration. The proposed method was able to decompose the progressive dopaminergic changes in the putamen into two orthogonal temporal progression curves associated with distinct spatial patterns.  This can be leveraged to assist in uncovering different mechanisms underlying the disease progression and disease initiation, or the sub-regions involved at different disease stages.  Xi \textit{et al.}~\cite{xi2019correlating} investigated the use of four different eigenmode algorithms:  POD, PCA, DMD and DMD with control (DMDC) as feature extracting techniques with the intention of enhancing the performance of a classifier to diagnose obstructive lung diseases using exhaled aerosol images. Their framework consisted of 3 phases. First, the researchers generated a database consisting of 405 exhaled aerosol images with physiology-based modeling and simulations. Second, feature extraction from the aerosol images was realized using the four mentioned algorithms. Finally, the extracted features were utilized by two classification techniques. Comparing the results, it was concluded that dynamic feature extractions (DMD and DMDC) significantly outperformed static algorithms (POD and PCA) achieving 94.8\% classification accuracy (see also~\cite{proctor2015discovering}).\\

In this work, we explore a new method that, to the authors knowledge has not yet been used in the field of medical imaging before: the Higher Order Dynamic Mode Decomposition (HODMD) technique~\cite{le2017higher}. As has been demonstrated in (\cite{le2017higher-1},~\cite{le2019alternative},~\cite{10.1007/978-3-030-20055-8_53},~\cite{le2019prediction}), the HODMD method is a more robust and accurate version of classical DMD, suitable for the analysis of complex signals and non-linear dynamical systems. The enhanced performance of HODMD, solidly grounded on mathematics and computer science principles, justifies our interest in its application for the analysis of medical images. The method indeed has sound foundations on the Koopman operator theory \cite{vega2020higher} and can be formulated as a fully data-driven technique. The HODMD method excels at identifying complex dynamics while presenting robust noise filtration properties. In this paper, HODMD was applied to the analysis of echocardiography images as a feature detection technique, with the aim at identifying and classifying cardiac diseases.  More specifically, six echocaerdiography datasets were analyzed, including one echocardiography taken from a healthy individual and five taken from  individuals that have been diagnosed with different cardiac diseases. The excellent capabilities of HODMD to identify and classify patterns in the medical images analyzed, introduces this technique as an automatic, robust and efficient tool, suitable for pathologies detection.  \\

The remainder of the paper will be organized as follows: in S (\ref{Methodology}) we will present the HODMD algorithm and its extension, the multidimensional HODMD, which are the algorithms used on this work. In S~(\ref{Medical images}) we present the data analyzed, together with a short explanation of the cardiac diseases investigated in this paper. S~(\ref{Results}) shows the results derived from the application of HODMD on the different datasets. Finally, S~(\ref{Con}) gathers the conclusions attained and discusses possible avenues to extend this work.

\section{Methodology}\label{Methodology}
\begin{figure*}[h!]

	\centering
	{\includegraphics[width=17cm, height=8cm]{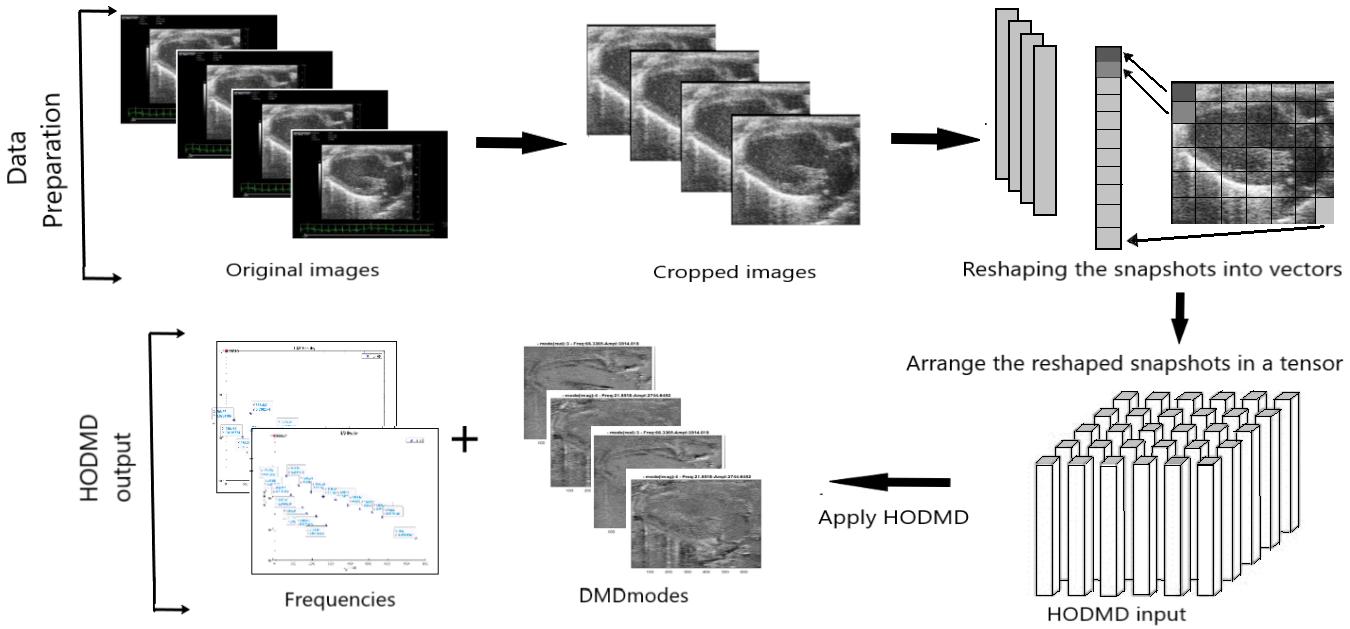}} \\

\centering
	\caption{Schematic diagram for the higher order dynamic mode decomposition (HODMD) analysis pipeline. The first step is data preparation: each frame (\textit{snapshot}) extracted from the video loop is cropped (removing the parts with the medical information), then reshaped into column vectors and arranged in a tensor, which is then used as the HODMD input. In the following steps, HODMD decomposes the reshaped data into a set of DMD modes, each associated with its own frequency, growth rate and amplitude.}	
	\label{Sket}
\end{figure*}

In this section we present \textit{Higher Order Dynamic Mode Decomposition} (HODMD)~\cite{le2017higher}, which is the method introduced in this work. Figure~\ref{Sket} presents a sketch describing the methodology of HODMD applied to analyze the medical images. Details about the algorithm are presented below.\\
HODMD is an extension of the well known technique, in the field of fluid dynamics, dynamic mode decomposition (DMD)~\cite{schmid2010dynamic}, generally used for the analysis of complex data modeling non-linear dynamical systems, solving different applications (e.g,~\cite{le2019alternative},~\cite{10.1007/978-3-030-20055-8_53},~\cite{le2019prediction}). Similarly to DMD, HODMD decomposes spatio-temporal data into a number of modes, each mode  related to a frequency, growth rate and amplitude, as presented in the following DMD expansion 
\begin{equation} \label{Eq0001}
\resizebox{.9\hsize}{!}{$\bm{v}(t)\simeq \sum\limits_{m=1}^M a_m\bm{u}_me^{(\delta_m+i\omega_m)(t-t_1)} \quad \textrm{for} \quad t_1\leq t \leq t_1+T ,$} 
\end{equation}
where $M$ is the number of DMD modes, $t$ is the time and $T$ is the sampled timespan, $\bm{u}_m$ are the normalized spatial modes, $a_m$ are the (real) amplitudes, and $\omega_m$ and $\delta_m$ are the associated frequencies and growth rates, respectively. The modes $\bm{u}_m$ represent the main patterns describing the data analyzed. Furthermore, one of the strong points of this algorithm is that the method also cleans noisy data and get rid of the frequencies with small amplitude, which represent irrelevant patterns, or simply noisy artifacts. \\

Before going into the details of the algorithm, we will briefly introduce the singular value decomposition, the higher order singular value decomposition algorithms and the Koopman operator, since they are used as part of the methodology in the HODMD algorithm. The Matlab codes of these algorithms and details about their multiple applications can be found in\cite{vega2020higher}.
\subsection{Preliminaries}

The following methods that we are introducing below are fully data-driven. For simplicity, the data is organized in matrix form in the following snapshot matrix :
\begin{equation} \label{Eq01}
\bm{ V}_1^K=[\bm{v}_1,\bm{v}_2,\dots , \bm{v}_K], 
\end{equation}

where $\bm{v_k}$ is a snapshot collected at time $t_k$ , with $ k = 1,\dots , K $. In the present article, each snapshot is a vector that contains the pixels of the image analyzed, hence $\bm{V}_1^K \in \mathbb{R}^{J\times K}$ (with $J=$ number of pixels in $X \times$ number of pixels in $Y$) .\\
The data can also be organized in tensor form, defined in discrete form as 
\begin{equation}\label{Eq001}
\resizebox{.9\hsize}{!}{$\bm{T}_{i_1,i_2,k}  \quad \textrm{for}\quad  i_1=1,\dots ,I_1  ;  i_2=1,\dots ,I_2 \quad \textrm{and}\quad  k=1,\dots ,K ,$}
\end{equation}
where $ i_1 $ and $ i_2 $ represent the position of each pixel in the plane containing the image, and $K$ is the number of snapshots.
 
\subsubsection{Singular value decomposition} 
Singular value decomposition (SVD) (\cite{lumley1967structure},~\cite{sirovich1987turbulence}) is a very powerful matrix decomposition tool and it is considered as one of the most important algorithms from the past decades.\\
Starting from the snapshot matrix eq. (\ref{Eq01}) the SVD will allow us to represent the matrix $\bm{V}_1^K$ as a product of three other matrices as follows:
\begin{equation}\label{SVD}
\bm{V}_1^K \simeq \bm{W} \bm{\Sigma}\bm{T}^T,
\end{equation}
where $\bm{W}$ $\in \bm{\mathbb{ C}}^{J\times J}$ and $\bm{T}$ $\bm{\in \mathbb{C}}^{K\times K}$ are unitary matrices and $J$ is number of SVD modes. The columns of $\bm{W}$ are called left singular vectors of $\bm{V}_1^K$ (related to spatial properties and orthogonal), the columns of $\bm{T}$ are called right singular vectors of $\bm{V}_1^K$ (related to temporal properties and orthogonal) and $\bm{\Sigma \in \mathbb{R}}^{J\times K}$ is a matrix with real, non negative entries on the diagonal and zeros off the diagonal, the elements of $\bm{\Sigma}$ are the singular values corresponding to the left and right singular vectors of $\bm{V}_1^K$ (see more details about the SVD algorithm in \cite{golub_singular_nodate}).

\subsubsection{Higher order singular value decomposition (HOSVD)} 
\begin{figure}[h!]

	\centering
	{\includegraphics[width=8cm, height=4cm]{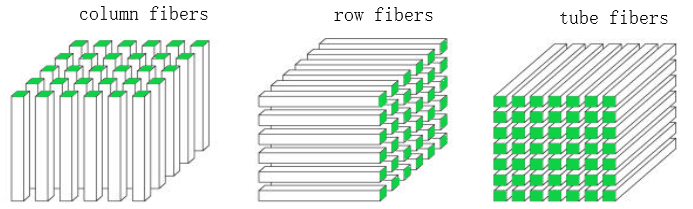}} \\

\centering
	\caption{Visualization of a third order tensor and its fibers.}	
	\label{Tens}
\end{figure}

\textit{Higher order singular value decomposition} (HOSVD) is a very robust extension of the SVD. It was originally introduced in 1966 by Tucker~\cite{tucker_mathematical_1966} and it resurfaced and extended in the use of several applications after the work presented by Lathauwer \textit{et al.}~\cite{doi:10.1137/S0895479896305696}. This method was developed to treat multidimensional databases called as Tensors.\\
HOSVD decomposes the tensor $\bm{T}$, eq. (\ref{Eq001}), as follows:
\begin{equation}
\resizebox{.9\hsize}{!}{$\bm{ T}_{i_1,i_2,... ,i_N}=\sum\limits_{n_1=1}^{r'_{1}}\sum\limits_{n_2=1}^{r'_2} ... \sum\limits_{n_N=1}^{r'_N} \bm{S}_{n_1n_2 ... n_N}\bm{U}^{1}_{i_1n_1}\bm{U}^2_{i_2n_2} ... \bm{U}^N_{i_Nn_N} ,$}
\end{equation}
such that, $ i_1=1,2 \dots ,I_1 $ ; $ i_2=1,2, \dots , I_2 $ ; $ \dots $ ; and $ i_n=1,2, \dots I_N $ . Denoting $ r_1 , r_2 , \dots \ r_N $ the ranks of the fibers of the tensor along the different dimensions, as presented in Fig. \ref{Tens}.
Thus $ r'_1 , \dots r'_N $ are such that $r'_1\geq max \{r_1,I_1\}, \dots , r'_N\geq max \{r_N,I_N \} $. The elements of the matrices $\bm{U}^1 ,\bm{U}^2, ... ,\bm{U}^N $ of sizes $ I_1\times r'_1 $ , $ I_2\times r'_2 $ , \dots $ I_N\times r'_N $ , respectively, are called \textit{mode matrices} and $\bm{S}_{n_1n_2\dots n_N} $ are the components of a $Nth$ order tensor $\bm{S}$ of size $r'_1\times r'_2 \times \dots \times r'_N $, known as the \textit{core tensor}.
 Furthermore, since the right side of the previous expression is considered as a \textit{tensor product} (denoted as \textit{tprod} below) of the core tensor and the \textit{mode matrices}, this could be written as follows: 
 \begin{equation}
\bm{T}= tprod(\bm{S}, \bm{U}^1 , \bm{U}^2 ,\dots , \bm{U}^N).
\end{equation}
\subsubsection{The \textit{Koopman} operator}
The \textit{Koopman} operator is a linear, infinite-dimensional operator that was introduced by Koopman in 1931~\cite{koopman1931hamiltonian}. It represents the action of a nonlinear dynamical system on the Hilbert space of measurement functions of the state. Mathematically, for a measurement function   $ g : \mathcal{D} \longrightarrow \mathbb{R} $ of the state $\bm{x}$, where $\mathcal{D}$ is an $n-$dimensional manifold.  The action of \textit{Koopman} operator is equal to the composition of the function $g$ with the flow map $\bm{F}$ as
\begin{equation}
 Kg= g\circ \bm{F}
\end{equation}
\begin{equation}
\Rightarrow Kg(\bm{x}_k)=g(\bm{F}(\bm{x}_k))=g(\bm{x}_{k+1}) ,
\end{equation}
So basically the \textit{Koopman} operator advances the measurements one time step into the future and remeasures the system at that next time step.
DMD algorithm uses the \textit{Koopman} operator in the discrete space, assuming that $g(\bm{x})$ is the identity.
\subsection{Higher  order dynamic mode decomposition}

As we stated before, HODMD is an extension of DMD, which relies on following  \textit{Koopman assumption}:
\begin{equation} \label{Eq2}
 \bm{v}_{k+1}= \bm{R}\bm{v}_k , \quad \textrm{for} \quad k= 1,..., K-1 , 
 \end{equation}
where $\bm{v}_k$ are our spatio-temporal data (the images from the ecocardiography) organized in $K$ equispaced $J$-dimensional \textit{snapshots} and $\bm{R}$ is \textit{the Koopman operator} .\\
Meanwhile, HODMD relies on a \textit{higher order Koopman assumption}:
\begin{equation} \label{Eq3}
\resizebox{.9\hsize}{!}{$\bm{v}_{k+d} = \bm{R}_1\bm{v}_k+\bm{R}_2\bm{v}_{k+1}+..+\bm{R}_d\bm{v}_{k+d-1} \hspace{0.1cm} \textrm{for} \hspace{0.1cm} k=1,.., K-d ,$}\hspace{-.2em}
\end{equation}
which relates $d$ subsequent snapshots. Note that when $d=1$, HODMD is equivalent to DMD.\\
Similarly to DMD this algorithm represents the spatio-temporal data $\bm{v}_k$ as an expansion of $M$ modes $\bm{u}_m$, each mode has his own amplitude $a_m$ , frequency $ \omega_m$ and growth rate $ \delta_m$ as follows :
\begin{equation} \label{Eq4}
\resizebox{.9\hsize}{!}{$ \bm{v}(t)\simeq \sum\limits_{m=1}^M a_m\bm{u}_me^{(\delta_m+i\omega_m)(t-t_1)} \quad \textrm{for} \quad t_1\leq t \leq t_1+T .$} \hspace{-.2em}
\end{equation}
HODMD algorithm can be summarized in three main steps.\\
\textbf{1-} Dimensionality reduction by applying the SVD (as in eq. (\ref{SVD})) to the full snapshot matrix eq. (\ref{Eq01}) as 
\begin{equation} \label{Eq5}
\bm{V}_1^K \simeq \bm{W} \bm{\Sigma}\bm{T}^T,
\end{equation}
where the number of retained modes $N$, is defined as \quad $ \sigma_{N+1}/\sigma_1 \leqslant \varepsilon_{SVD} $, where
$\sigma_1,\dots , \sigma_N $ are the singular values and the threshold $\varepsilon_{SVD} $ is selected according to the level of noise in the data.\\
The previous equation can be written as :
\begin{equation} \label{Eq6}
\bm{ V}^K_1 \simeq  \bm{W} \bm{\widehat{V}}_1^K , \quad \textrm{where} \quad  \bm{\widehat{V}}^K_1 = \bm{\Sigma} \bm{T}^T , 
\end{equation}
$\bm{\widehat{V}}^K_1 $ will be called \textit{the reduced snapshot matrix}.\\
\textbf{2-} In the second step, we use the \textit{higher order Koopman assumption} defined in eq. (\ref{Eq3}) to the reduced snapshot matrix as 
\begin{equation} \label{Eq7}
 \bm{\widehat{V}}^K_{d+1}\simeq \bm{\widehat{R}}_1\bm{\widehat{V}}^{K-d}_1+\bm{\widehat{R}}_2\bm{\widehat{V}}^{K-d+1}_2 + \dots + \bm{\widehat{R}}_d\bm{\widehat{V}}^{K-1}_d ,  
\end{equation}
where $ \bm{ \widehat{R}}_k=\bm{ W}^T \bm{R}_k \bm{W} $.\\
This equation can be represented using the reduced snapshot matrix and the modified \textit{Koopman} matrix $ \bm {\tilde{R}}$ as follows 
\begin{equation} \label{Eq8}
 \bm{\widetilde{V}}_2^{K-d+1} =\bm{ \widetilde{R}}\bm{\widetilde{V}}_1^{K-d} ,   
\end{equation}
where 
$$
\bm{\tilde{V}}_1^{K-d}= \begin{bmatrix} 
\bm{\widehat{V}}_1^{K-d} \\
\bm{\widehat{V}}_2^{K-d+1} \\
\dots \\
\bm{\widehat{V}}_d^{K-1}  
\end{bmatrix} ,\quad 
\bm{\tilde{V}}_2^{K-d+1}= \begin{bmatrix} 
\bm{\widehat{V}}_2^{K-d+1} \\
\dots \\
\bm{\widehat{V}}_d^{K-1} \\
\bm{\widehat{V}}_{d+1}^K  
\end{bmatrix} , $$
$$
\bm{\tilde{R}}=\begin{bmatrix} 
\bm{0} & \bm{I} & \bm{0} & \dots & \bm{0} & \bm{0} \\
\bm{0} & \bm{0} & \bm{I} & \dots &\bm{0} & \bm{0} \\
\dots & \dots & \dots & \dots & \dots & \dots \\
\bm{0} & \bm{0} & \bm{0} & \dots & \bm{I} & \bm{0} \\
\bm{\widehat{R}}_1 & \bm{\widehat{R}}_2 & \bm{\widehat{R}}_3 & \dots &\bm{ \widehat{R}}_{d-1} & \bm{\widehat{R}}_d    
\end{bmatrix}.
$$
A second dimensionality reduction is carried out to the matrix containing the reduced snapshots using SVD and the tolerance $ \varepsilon_{SVD}$ as \quad $\tilde{\sigma}_{N'+1}/\tilde{\sigma}_1< \varepsilon_{SVD} $, where $N'$ is the number of retained SVD modes and $ \tilde{\sigma}_i $ are the singular values. This truncation yields 
\begin{equation} \label{Eq9}
\resizebox{.9\hsize}{!}{$ \bm{ \tilde{V}}_1^{K-d+1} \simeq \bm{\tilde{U}} \bm{\tilde{\Sigma}} \bm{\tilde{T}}^T \simeq  \bm{\tilde{U}} \bm{\overline{T}}^{K-d+1}_1 ,\textrm{with} \quad \bm{\overline{T}}^{K-d+1}_1 = \bm{\tilde{\Sigma}} \bm{\tilde{T}}^T ,$}\hspace{-.2em}
\end{equation}
this step is completed through pre-multiplying eq. (\ref{Eq8}) by $\bm{\tilde{U}}^T$, and invoking eq. (\ref{Eq9}) it gives : 
\begin{equation} \label{Eq10}
 \bm{ \overline{T}}_2^{K-d+1} = \bm{\bar{R}} \bm{\overline{T}}_1^{K-d} , 
\end{equation}
such that $\bm{\overline{R}} \in N'\times N'$  is the new \textit{Koopman matrix} defined as $\bm{ \overline{R}} =\bm{ \tilde{U}}^T \bm{\tilde{R}} \bm{\tilde{U}}$,  but we are not computing $\bm{\overline{R}}$ with this expression, instead we use the methodology presented in the next step.\\
\textbf{3-} The third step is computing the DMD modes, frequencies and growth rates, in order to do that, $\bm{\overline{R}}$ must be computed first, which is simply done by applying SVD on the matrix  $ \bm{\overline{T}}_1^{K-d} $  
\begin{equation} \label{Eq11}
 \bm{ \overline{T}}_1^{K-d} =\bm{ U} \bm{\Lambda}\bm{ V}^T ,
\end{equation}
and then we substitute eq. (\ref{Eq11}) in eq. (\ref{Eq10}) and multiply the result by $ \bm{V} \bm{\Lambda}^{-1} \bm{U}^T $ to obtain :
\begin{equation}\label{Eq12}
 \bm{ \overline{R}} =\bm{ \overline{T}}_2^{K-d+1}\bm{V} \bm{\Lambda}^{-1}\bm{ U}^T ,  
\end{equation}
once the matrix $\bm{\overline{R}}$ has been calculated, the reduced DMD 
expansion for the reduced snapshots eq. (\ref{Eq6})~ can be computed as follows 
\begin{equation} \label{Eq13}
  \bm{\hat{v}}_k = \sum\limits_{m=1}^{M} \hat{a}_m \bm{\hat{u}}_m e^{(\delta_m+i\omega_m)t_k} ,\quad  \textrm{for} \quad  k= 1, \dots ,K ,
\end{equation}
the reduced DMD modes $ \bm{\hat{u}}_m$ were  calculated by keeping the first $ M $ elements of the vector $ \bm{\hat{q}}_m =\bm{ \tilde{U}} \bm{\bar{q}}_m $ , where $ \bm{\bar{q}}_m$ represents the eigenvectors of $\bm{ \overline{R}}$ and the associated eigenvalues $ \mu_m  $ provides the frequencies $\omega_m$ and growth rates $ \delta_m$ by the following expression:
 \begin{equation} \label{Eq14}
\delta_m + i\omega_m = log( \mu_m)/\Delta t .
\end{equation}
\textbf{4-}  The fourth and final step is computing the amplitudes $ a_m $ via least-square fitting of eq. (\ref{Eq13}).
Finally, the DMD modes are ordered with respect to there amplitudes in a decreasing order. It is possible to determine the number of $M$ modes to retrain in the DMD expansion eq. (\ref{Eq4}) using a different tolerance $ \varepsilon_{DMD} $  such as \quad $ a_{M+1}/a_1 \leqslant \varepsilon_{DMD} $\quad and finally computing the DMD expansion for the original \textit{snapshots}.\\
The error of the HODMD reconstruction  eq. (\ref{Eq4}) is measured using the relative root mean square error (RRMSE) as follows
\begin{equation}\label{RMS}
\textrm{RRMSE} = \sqrt{\dfrac{\sum_{k=1}^{K}\lVert \bm{v}_k - \bm{v}_k^{DMD}\rVert^2_2}{\sum_{k=1}^K \lVert \bm{v}_k \rVert^2_2}} ,
\end{equation}
where $\bm{v}_k$ are the original snapshots, $\bm{v}_k^{DMD}$ are their reconstruction using the HODMD algorithm and $\lVert . \rVert_2$ is the Euclidean norm. 
\subsection{Multidimensional higher order dynamic mode decomposition}\label{multidimensional HODMD}
\textit{Multi-dimensional iterative HODMD} is an extension of HODMD, introduced in~\cite{le2017higher-1}  
for the analysis and pattern identification in complex experimental data, highly noisy and non-linear dynamics. This algorithm organizes the data in tensor form as in eq. (\ref{Eq001}) and replaces the SVD algorithm by the HOSVD algorithm in HODMD. In this way, it is possible to better clean the data, since HOSVD applies SVD along each spatial direction. The iterative algorithm simply applies the multi-dimensional HODMD iteratively. In other words, HODMD obtains the extension eq. (\ref{Eq4}), then the method is applied iteratively over this data reconstruction until the number of HOSVD modes is the same between two consecutive iterations.
The algorithm has been validated and successfully tested in several applications (see~\cite{le2019alternative},~\cite{10.1007/978-3-030-20055-8_53},~\cite{le2019prediction}). More details about the algorithm can be found in~\cite{le2017higher-1}.\\
Unlike standard HODMD where the data is organized in a snapshot matrix, in this case the data matrix is substituted by a multidimensional snapshot \textit{tensor} and the expansion eq. (\ref{Eq4}) in discrete and tensorial form is defined as follows
\begin{equation}
\resizebox{.9\hsize}{!}{$\bm{T}_{j_1j_2k} \simeq \sum\limits_{m=1}^{M} a_m\bm{U}_{j_1j_2m}e^{(\delta_m+i\omega_m)(k-1)\Delta t}  ,\quad \textrm{for} \quad k = 1,\dots ,K , $}\hspace{-.2em}
\end{equation}
where\quad $ \boldmath{j_1 = 1,\dots ,J_1 ; j_2 = 1,\dots ,J_2} $ ( $J_k$ is the size of the vector $ T_{j_1j_2k} $ ).\\
This algorithm has two main steps, the first one is a dimensionality reduction, but instead of using SVD, HOSVD will be applied to \textit{the snapshot tensor}, which yields the following decomposition :
\begin{equation}
 \bm{T}_{j_1j_2k}\simeq \sum\limits_{p1=1}^{P_1}\sum\limits_{p_2=1}^{P_2}\sum\limits_{n=1}^{N} \bm{S}_{p_1p_2n}\bm{W}^{(1)}_{j_1p_1}\bm{W}^{(2)}_{j_2p_2}\bm{\mathsf{T}}_{kn}, 
\end{equation}

where $ \bm{S} $ is the \textit{core tensor}, the columns of $ \bm{W}^{(1)} , \bm{W}^{(2)} $ are called the spatial \textit{ modes}, and the columns  of $ \bm{\mathsf{T}} $ are called the temporal \textit{modes} of the decomposition.\\
With respect to a spatial tolerance and a temporal tolerance we determine the number of modes to retrain from each one of the \textit{modes}. And finally steps from 2 to 4 of the standard HODMD are applied to the  temporal modes $ \bm{\mathsf{T}} $.
\section{Datasets description: the echocardiography images}\label{Medical images}
This section introduces the datasets analyzed, which consist of several medical images from an ecocardiography. The database includes images associated to the following cardiac pathologies in the heart:
\begin{itemize}
\item Diabetic cardiomyopathy, which is a disorder characterized by structural remodeling in the myocardium in people with diabetes mellitus. It can lead to inability of the heart to circulate blood through the body effectively. Due to various possible causes, blood moves through the heart and body at a slower rate, and pressure in cardiac chambers in the heart increases.
\item Obesity, it is a complex disease involving an excessive amount of body fat, the more fat on the body, the greater the strain on the heart. Obesity forces the heart to pump harder to distribute the blood throughout the body. This in turn increases the chance of developing rapid heart rate.
\item Cardiac hypertrophy (TAC hypertrophy, SFSR4), which is the abnormal enlargement, or thickening of the heart muscle. Thickened heart muscle, can cause changes in cardiac motion, the heart's electrical system, resulting in fast or irregular heartbeats. Causes of this disease can be physiological – for example, the amount of exercise performed by an athlete – or pathological – for example, as a result of hypertension or valvular disease. 
\item Myocardial infarction, commonly known as a heart attack. It occurs when blood flow decreases or stops to a part of the heart, causing damage to the heart muscle. The usual cause of sudden blockage in a coronary artery is the formation of a blood clot (thrombus) causing the heart muscle to becomes "starved" for oxygen, and consequently, causing a permanent damage.
 \end{itemize}
\subsection{Material}
All the images have been obtained from previous mouse model of cardiac diseases performed in accordance with protocols approved by the \textit{Centro Nacional de Investigaciones Cardiovasculares} (CNIC) Institutional Animal Care and Research Advisory Committee of the Ethics Committee of the Regional Government of Madrid (PROEX177/17).\\

\textbf{Echocardiography images acquisition }\\  
Transthoracic echocardiography was performed under isoflurane anesthesia by an expert operator using a high-frequency ultrasound system (Vevo 2100, Visualsonics Inc, Canada) with a 40-MHz linear probe. Isoflurane was administered in 100\% oxygen and the dose was adjusted to maintain podal reflex (light anesthesia plane). Mice were placed in supine position using a heating platform and warmed ultrasound gel was provided to preserve normothermia. A base apex electrocardiogram was used for heart rate and rhythm continuous monitoring. Standard bidimensional (2D) parasternal long and short axis views (LAX and SAX, respectively) of the left ventricle (LV) were obtained as previously described~\cite{villalba2017lung}.\\
Offline, LAX and SAX video loops from different mouse model of cardiac diseases (diabetes, obesity, hypertrophy and infarction) and from healthy mice were exported as DICOM format, including at least 3 cardiac cycles.\\

\textbf{Mouse model diseases}\\
Hearts from healthy C57BL/6 10-weeks-old mice were used as control (CTL). Diabetes was induced in mice by injecting streptozotocin (STZ, 50mg/kg, 0.05mol/L in citrate buffer, pH 4.5, Sigma, St. Louis, USA) i.p for five consecutive days ~\cite{villalba2017lung} and images were assessed 16 weeks post-induction. Obesity was induced feeding mice with Western diet [45 kcal\% (24 g\%) palm oil-based fat, 35 kcal\% (41 g\%) carbohydrate, 20 kcal\% (24 g\%) protein; based on OpenSource Diets No. D12451, Research Diet Services, Wijk bij Duurstede, The Netherlands] and images were assessed 88 weeks post-induction. Cardiac hypertrophy images were obtained from mutant mice previously described (SFSR4 KO~\cite{larrasa2021srsf4}) and images were assessed 24 weeks of age. Another model of hypertrophy was induced using the aorta constriction surgery~\cite{lara2015guidelines} and images were assessed 4 weeks post-induction. Lastly, a model of myocardial infarction was used from mice subjected to left anterior descending coronary artery permanent ligation (~\cite{lara2015guidelines},~\cite{villalba2017systolic}) and images were assessed 4 weeks post-induction.    
\section{Results}\label{Results}
\begin{figure}[h!]
    \centering
    \includegraphics[width=9cm, height=6.5cm]{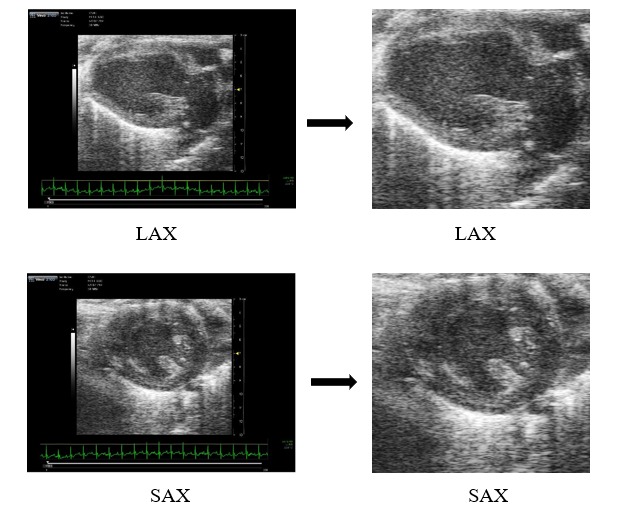}
    \caption{The figure shows the original images on the left (pre-cropping), and on the right the images after removing the parts containing the medical information (post-cropping).}
    \label{Proc}
\end{figure}

In this section, we detail how HODMD was used to analyze the medical images. As mentioned before, we initially analyzed six echocardiography datasets, where each dataset encompasses two video loops taken from LAX view and SAX view, with 200 to 300 snapshot per video.\\
 All the datasets went through the same procedure into the Matlab code in order to be analyzed using the HODMD algorithm. The data, which was in DICOM format was imported. All the frames from each video were first extracted and then cropped, parts of the image that contain medical details were removed, focusing only on the heart area in the image (see Fig. \ref{Proc}). The resulted images were converted to grayscale images, reshaped into vectors and arranged in a tensor (see Fig. \ref{Sket}).\\
HODMD requires the selection of several tunable parameters:
\begin{itemize}
\item The number of snapshots $K$ is first chosen, this selection is not critical and could be done approximately by trying to capture some of the full cardiac cycles.
\item Two thresholds, $ \varepsilon_{SVD} $ for the the dimension-reduction steps and $ \varepsilon_{DMD} $ for amplitude truncation. $ \varepsilon_{SVD} $ should be somewhat larger than the noise level, meanwhile, decreasing $ \varepsilon_{DMD} $ increases the number of the identified frequencies. 
\item The time step or the time between the snapshots $ \Delta t $ is estimated to be 4 milliseconds and the timespan $T$.
\item Finally, the index $d$, has been selected in a calibration process, starting from $ d\simeq K/10 $ and increasing or decreasing its value until we minimize the relative RRMSE (eq. (\ref{RMS})).
 \end{itemize}
After some calibration, setting the correct parameters to minimize the RRMS error (see details in\cite{vega2020higher}), HODMD was applied using the following parameters: the number of snapshots $K$ was set to 100 snapshot in all of the cases except for healthy LAX data, diabetic cardiomyopathy (both LAX and SAX) and obesity (SAX), such that, in these cases $K$ was set to 200 snapshots, where the noise level in these datasets obligated us to analyze a larger number of snapshots, such that, they cover a sufficient number of cardiac cycles allowing us to capture the relative frequencies in two parallel lines. The thresholds were fixed to be $  \varepsilon_{SVD} = \varepsilon_{DMD} = 5\times 10^{-4} $ for all the datasets. The time step is given by $ \Delta t = 4\times 10^{-3} $ , meanwhile, the timespan $ T\in [1,K] $ is scaled with the time step $ \Delta t $ in the cases where $ K=100 $ and with $ 2\Delta t $ in the cases where $ K = 200 $. Finally, the the index $d$ varies between (30 to 35) when $ K=100 $ and between (60 to 70) in the cases where $ K = 200 $, in good agreement with the calibration process described in \cite{le2017higher} ($d$ scales with the number of snapshots). 
\subsection{Healthy data}     

\begin{figure*}[h!]
	\centering
\textbf{Healthy heart results}\par\medskip	
	\subfloat[ Frequency obtained from the LAX view ]{\includegraphics[width=8.2cm, height=6cm]{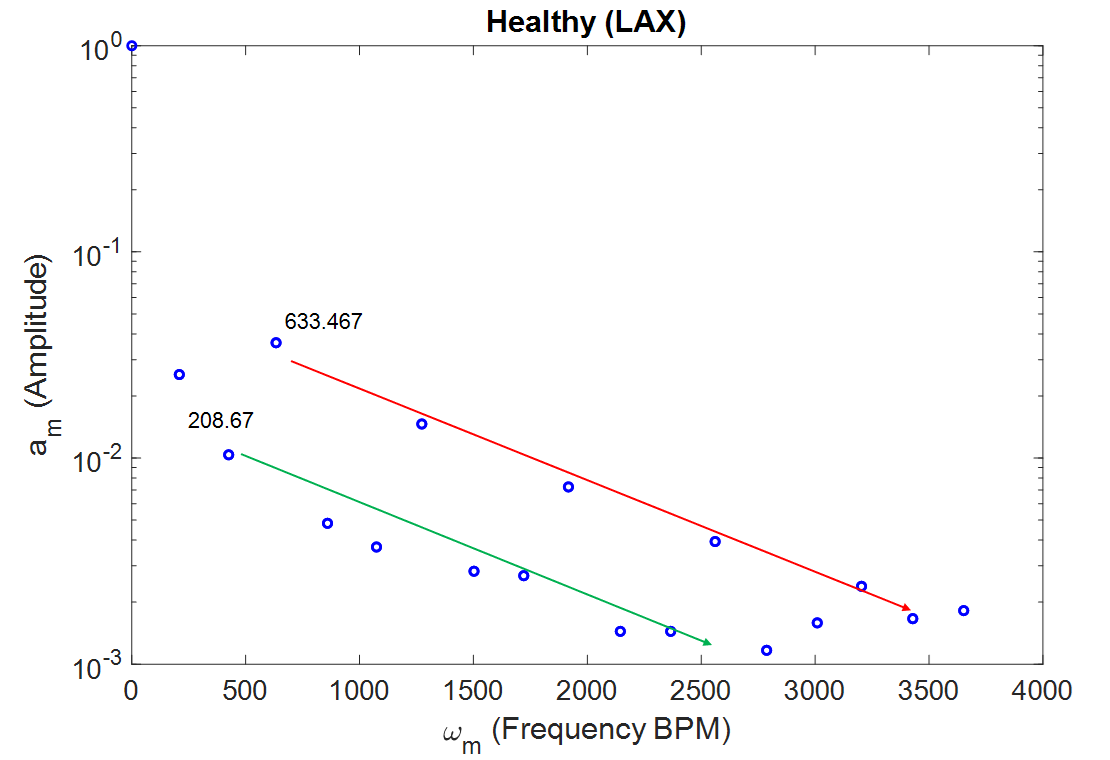}} \qquad 
\subfloat[ Frequency obtained from the SAX view ]{\includegraphics[width=8.2cm, height=6cm]{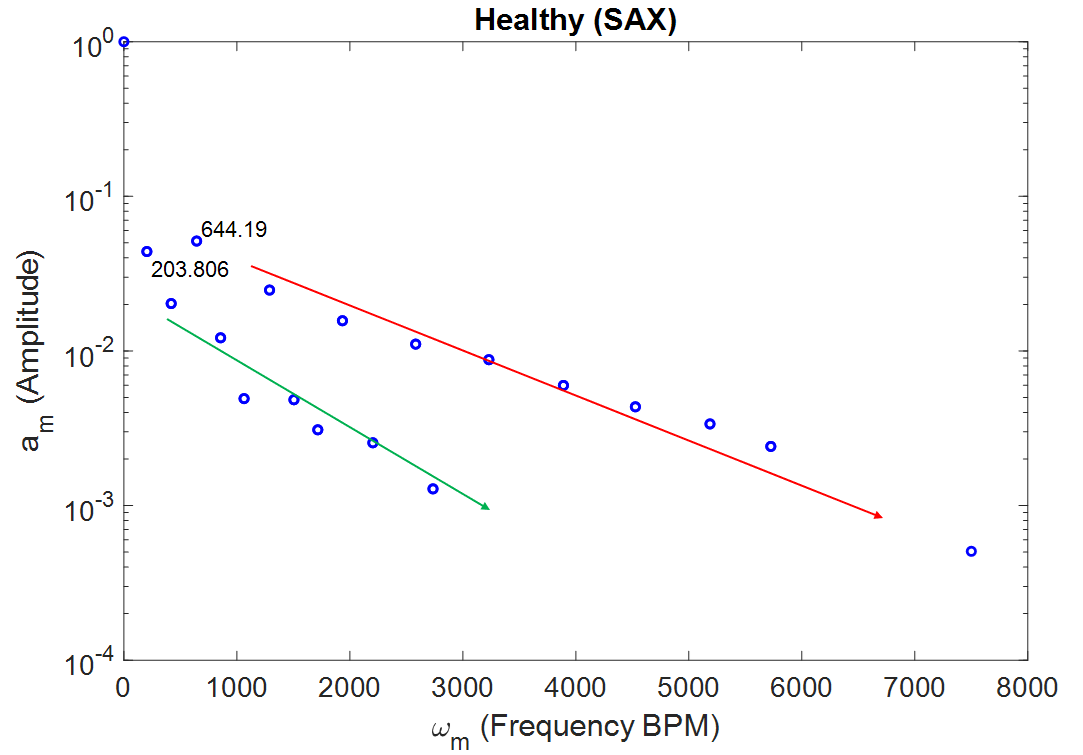}}\\
\hspace{-0.1cm}\subfloat[Original LAX view]{\includegraphics[width=4.4cm, height=3.5cm]{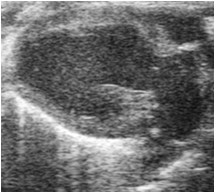}} \hspace{4.3cm} 
\subfloat[Original SAX view]{\includegraphics[width=4.4cm, height=3.5cm]{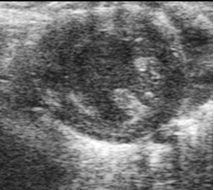}}\\
	\subfloat[ Upper branch mode ]{\includegraphics[width=4.4cm, height=3.5cm]{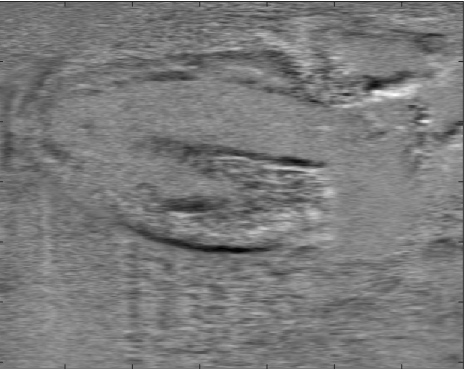}} \hspace{4.2cm} 
\subfloat[ Upper branch mode ]{\includegraphics[width=4.5cm, height=3.5cm]{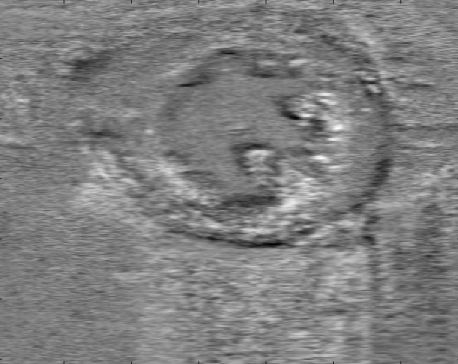}}\\	 
	\subfloat[Lower branch mode ]{\includegraphics[width=4.4cm, height=3.5cm]{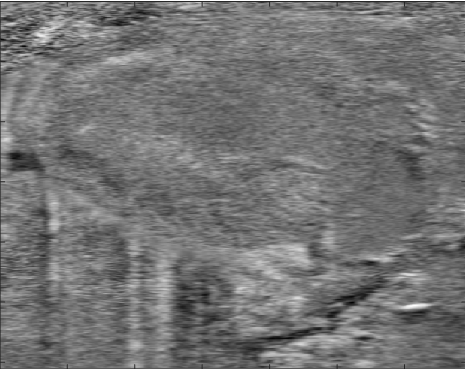}}	\hspace{4.2cm} 
	\subfloat[Lower branch mode ]{\includegraphics[width=4.5cm, height=3.5cm]{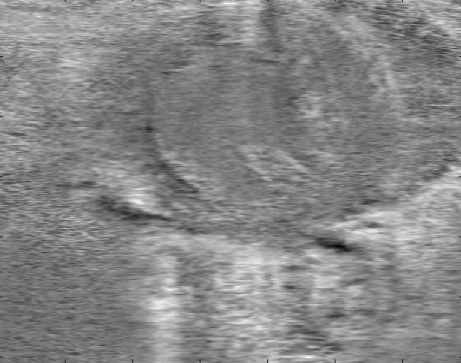}}		
\\
\centering
	\caption{The frequencies captured from analyzing the data of the healthy heart (LAX on the left and SAX on the right) and the dominant modes related to each branch of frequency. The data are normalized with their maximum value. The color-scale correspond to -1 (black), 0 (grey), 1 (white). the values of the frequencies from the upper and lower branches can be found in Table \ref{H_Freq}  in Appendix.}	
	\label{LAX_H2}
\end{figure*}

The first dataset that was analyzed by our algorithm was the healthy data. As it is shown in Fig. \ref{LAX_H2}, HODMD was able to capture two groups of signals representing two parallel lines, each one composed of a different group  of frequencies. These two groups of signals, which are periodic and regular, are related to the heart rate (upper branch) and the respiratory rate (lower branch). The dominant frequency in the upper branch is 633 beat per minute (BPM) and in the lower branch is 208 breath per minute (BPM) in the LAX results, and 644/BPM in the upper branch and  203/BPM in the lower branch in the SAX results (the values of the frequencies from the upper and lower branches can be found in Table \ref{H_Freq} in Appendix). Six and nine harmonics of these frequencies are identified in the upper and lower branches, respectively, for the LAX data. Meanwhile, ten and eight harmonics are identified in the upper and lower branches, respectively, for the SAX data. The DMD modes support these results, as seen in the bottom part of Fig. \ref{LAX_H2}. The modes from the upper branch present their highest intensity in the area representing the heart, and the ones from the lower branch present their higher intensity in the area of the image representing the lungs. 

\subsection{Unhealthy data}
By applying HODMD to the unhealthy echocardiography images, as it is seen in ( Figures~\ref{DB},~\ref{Hyper_Modes},~\ref{Hyper_Freq},~\ref{MI}) we still capture the two lines of frequencies in each case. Although, the frequencies might have some changes when compared to the healthy data, this fact is not  necessarily related to the disease since all the mice go through anesthesia to try to keep the heart rate stable. Also we note that there is a difference between the frequencies obtained from the LAX and the ones obtained from the SAX because in  the case of SAX view, the mice are receiving a higher doze of anesthesia. Nevertheless, the different patterns identified in the DMD modes, reveal the type of pathology studied. These modes are strong representers of the shape of the heart in certain unhealthy conditions or representers of certain features that are related to a certain disease. The results obtained from analyzing the pathological heart datasets can be sorted into two main categories: myocardial infarction model (represented in the results obtained from the myocardial infarction datasets) and hypertrophy models (represented in the results obtained from obesity, TAC hypertrophy and SFSR4 hypertrophy datasets ). Regarding the results from the Diabetic cardiomyopathy, the DMD modes did not capture any abnormalities or features representing the disease.\\ 
\newpage
\subsubsection{Case 01: Diabetic Cardiomyopathy}
\begin{figure*}[h]
\textbf{Case 01-Diabetic Cardiomyopathy}\par\medskip
	\centering
	
	\subfloat[ Frequency of the heart with Diabetic Cardiomyopathy (LAX) ]{\includegraphics[width=8.1cm, height=6cm]{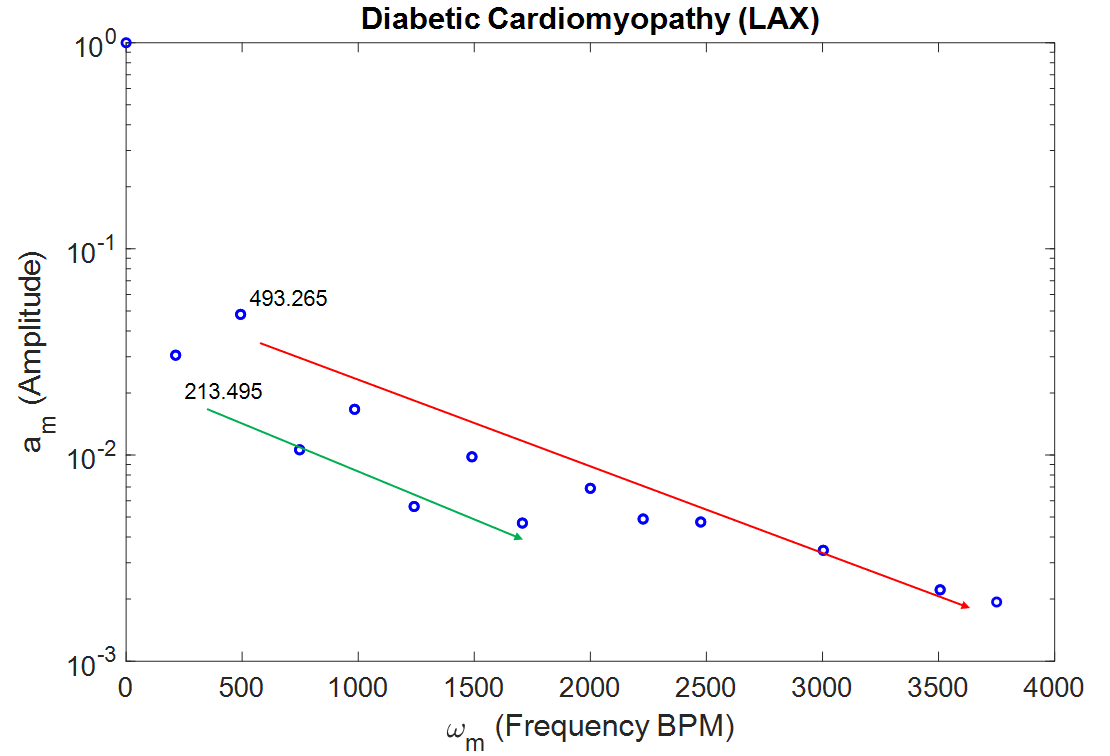}} \qquad
\subfloat[ Frequency of the heart with Diabetic Cardiomyopathy (SAX) ]{\includegraphics[width=8.1cm, height=6cm]{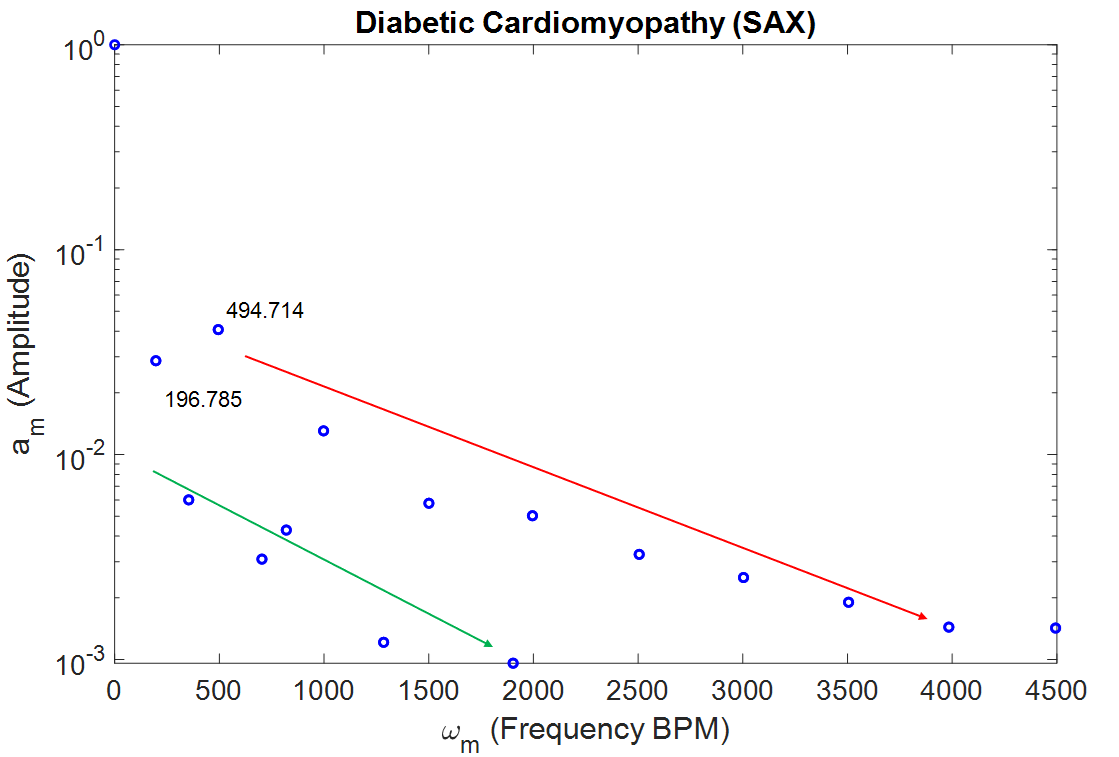}}\\
\subfloat[Original LAX view]{\includegraphics[width=4.4cm, height=3.5cm]{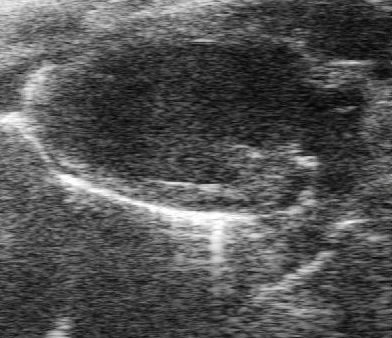}} \hspace{3.3cm}
\subfloat[Original SAX view]{\includegraphics[width=4.4cm, height=3.5cm]{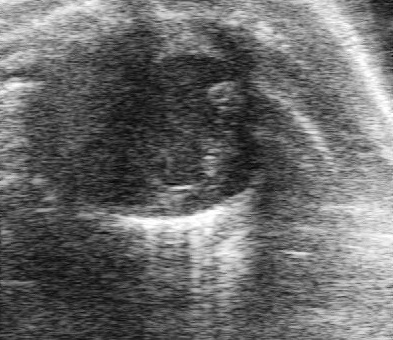}} \\
	\subfloat[ Upper branch mode ]{\includegraphics[width=4.4cm, height=3.5cm]{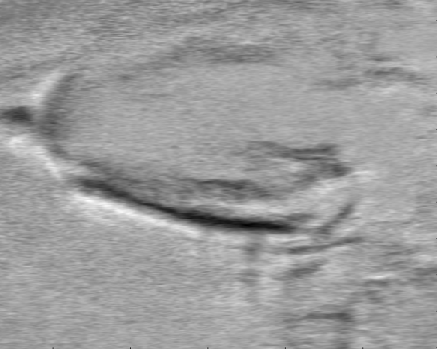}} \hspace{3.3cm}
\subfloat[ Upper branch mode ]{\includegraphics[width=4.5cm, height=3.5cm]{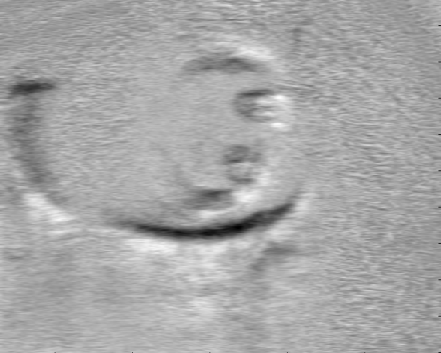}} \\	 	 
\subfloat[Lower branch mode ]{\includegraphics[width=4.4cm, height=3.5cm]{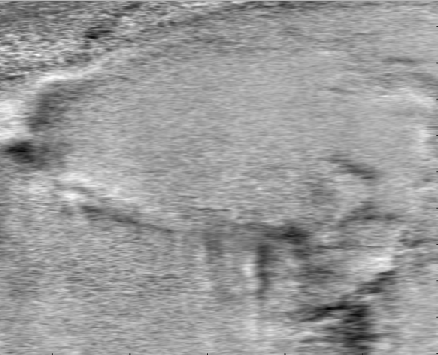}}	\hspace{3.3cm} 
	\subfloat[Lower branch mode ]{\includegraphics[width=4.5cm, height=3.5cm]{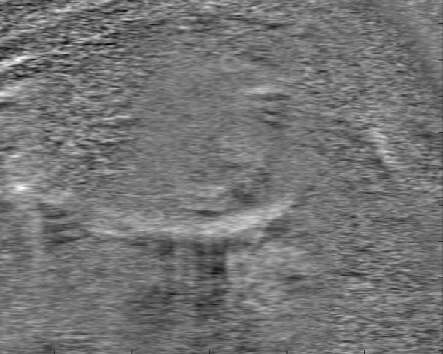}}		
\\
\centering
	\caption{The frequencies captured from analyzing the data of the hearts with Diabetic Cardiomyopathy and the dominant modes related to each branch of frequency. The data are normalized with their maximum value. The color-scale correspond to -1 (black), 0 (grey), 1 (white). The values of the frequencies from the upper and lower branches can be found in Table \ref{DC_Freq} in Appendix. }
	\label{DB}
\end{figure*}

 In the case of the Diabetic Cardiomyopathy presented in Fig. \ref{DB}, no abnormalities were captured. Although the echocardiography was taken from mice with diabetic cardiomyopathy, the echo does not capture any changes in the shape of the heart compared to the healthy dataset analyzed, as expected in this type of pathology. In this case it is worth to mention that there is a noticeable change in the heart rate and the respiratory rate (in SAX) compared to the frequencies identified in the healthy case (see Table \ref{DC_Freq} in Appendix). This change in the respiratory rate could be related to the effect of anesthesia. However, the frequency has noticeably decreased, which is something expected in the case of the present pathology (diabetic cardiomyopathy is related to a slow heart rate). HODMD reveals this change in the frequency, while the shape of the DMD is similar as the DMD modes in the healthy case, also as expected in this pathology.\\
\subsubsection{Case 02: Hypertension models}
\begin{figure*}[h!]
\textbf{Case 02- Hypertension models: }\par\medskip	
\textit{01- DMD modes:}\par\medskip	

	\centering

	\textbf{ LAX DMD modes \hspace{4.6cm} SAX  DMD modes }\par\medskip	
\rotatebox{90}{\hspace{0.7cm}\textbf{Healthy}}\hspace{0.1cm}	\subfloat[Upper branch] {\includegraphics[width=3.8cm, height=3.2cm]{H_LAX_up2}} \hspace{0.2cm} 
\subfloat[Lower branch ]{\includegraphics[width=3.8cm, height=3.2cm]{H_LAX_low2}}	\hspace{0.4cm} 	
\subfloat[ Upper branch ]{\includegraphics[width=3.8cm, height=3.2cm]{H_SAX_up2}} \hspace{0.2cm}
	\subfloat[Lower branch ]{\includegraphics[width=3.8cm, height=3.2cm]{H_SAX_low2}}		
\\
\rotatebox{90}{\hspace{0.7cm}\textbf{Obesity}}\hspace{0.1cm}		\subfloat[ Upper branch ]{\includegraphics[width=3.8cm, height=3.2cm]{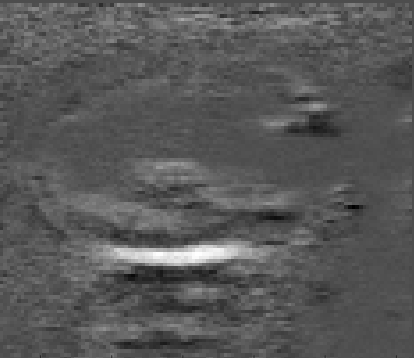}} \hspace{0.2cm}
\subfloat[Lower branch ]{\includegraphics[width=3.8cm, height=3.2cm]{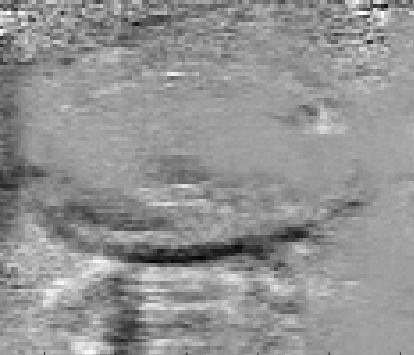}}	\hspace{0.4cm} 		
\subfloat[ Upper branch ]{\includegraphics[width=3.8cm, height=3.2cm]{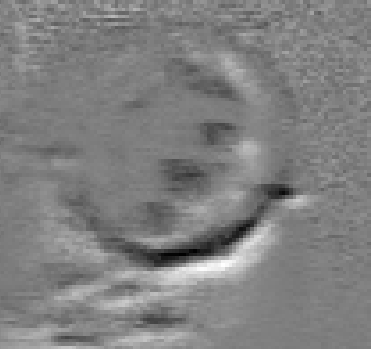}} 	\hspace{0.2cm}
	\subfloat[Lower branch]{\includegraphics[width=3.8cm, height=3.2cm]{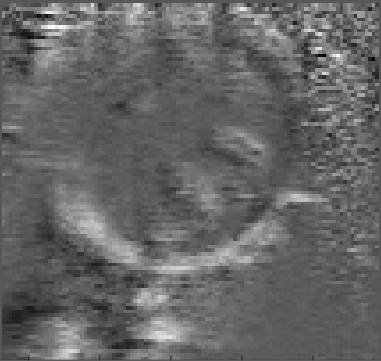}}		
\\
\rotatebox{90}{\textbf{TAC Hypertrophy}}\hspace{0.1cm}		\subfloat[ Upper branch]{\includegraphics[width=3.8cm, height=3.2cm]{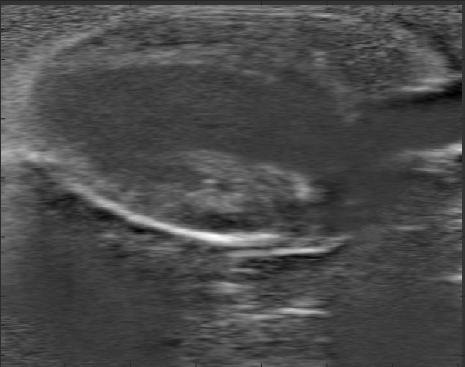}} \hspace{0.2cm} 
\subfloat[Lower branch]{\includegraphics[width=3.8cm, height=3.2cm]{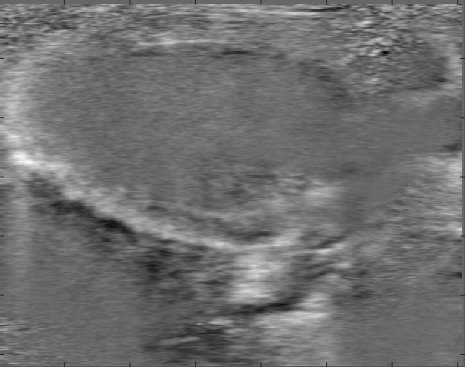}}	\hspace{0.4cm} 	
\subfloat[ Upper branch ]{\includegraphics[width=3.8cm, height=3.2cm]{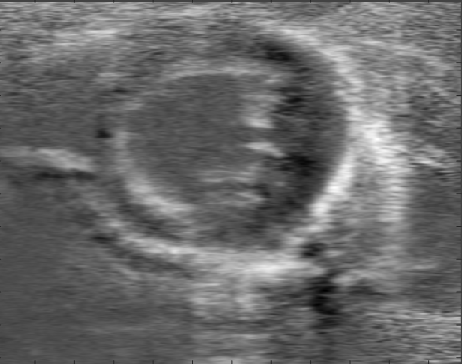}} \hspace{0.2cm}
	\subfloat[Lower branch]{\includegraphics[width=3.8cm, height=3.2cm]{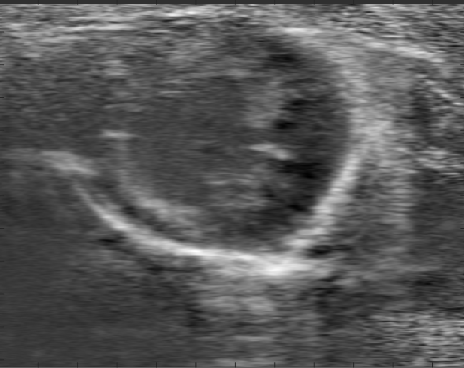}}		
\\
\rotatebox{90}{\textbf{SFSR4 Hypertrophy}}\hspace{0.1cm}		\subfloat[ Upper branch ]{\includegraphics[width=3.8cm, height=3.2cm]{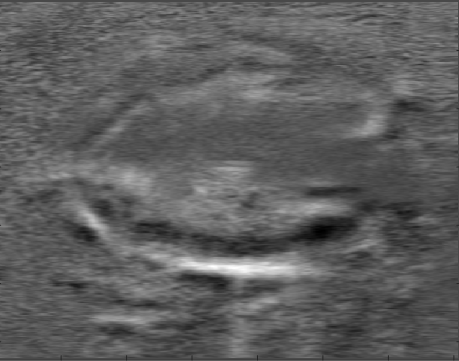}} \hspace{0.2cm}
\subfloat[Lower branch  ]{\includegraphics[width=3.8cm, height=3.2cm]{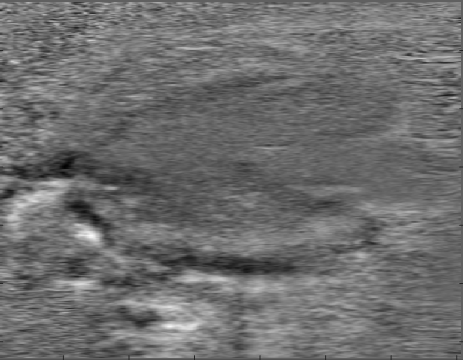}}	\hspace{0.4cm} 		
\subfloat[ Upper branch  ]{\includegraphics[width=3.8cm, height=3.2cm]{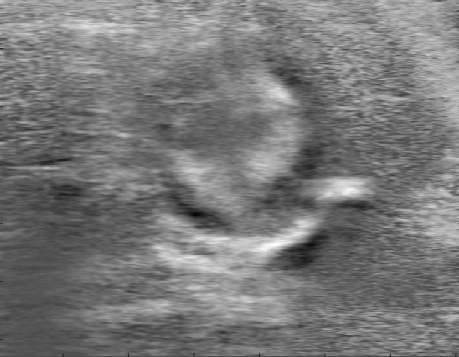}} 	\hspace{0.2cm}
	\subfloat[Lower branch  ]{\includegraphics[width=3.8cm, height=3.2cm]{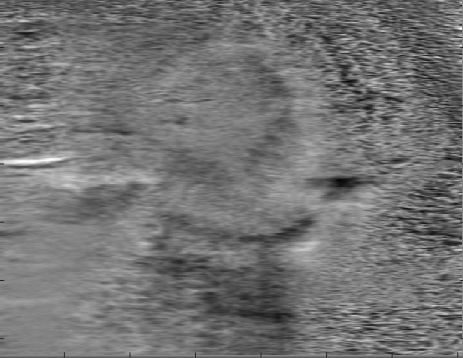}}	\\

\centering
	\caption{A comparison between the DMD modes obtained from the healthy data set and the DMD modes obtained from analyzing the datasets of the hypertrophic hearts.}	
	\label{Hyper_Modes}
\end{figure*}

\begin{figure*}[h!]
\textit{02- Frequencies:}\par\medskip
	\centering
	
\subfloat[ Frequency of the heart with Obesity (LAX) ]{\includegraphics[width=8.1cm, height=6cm]{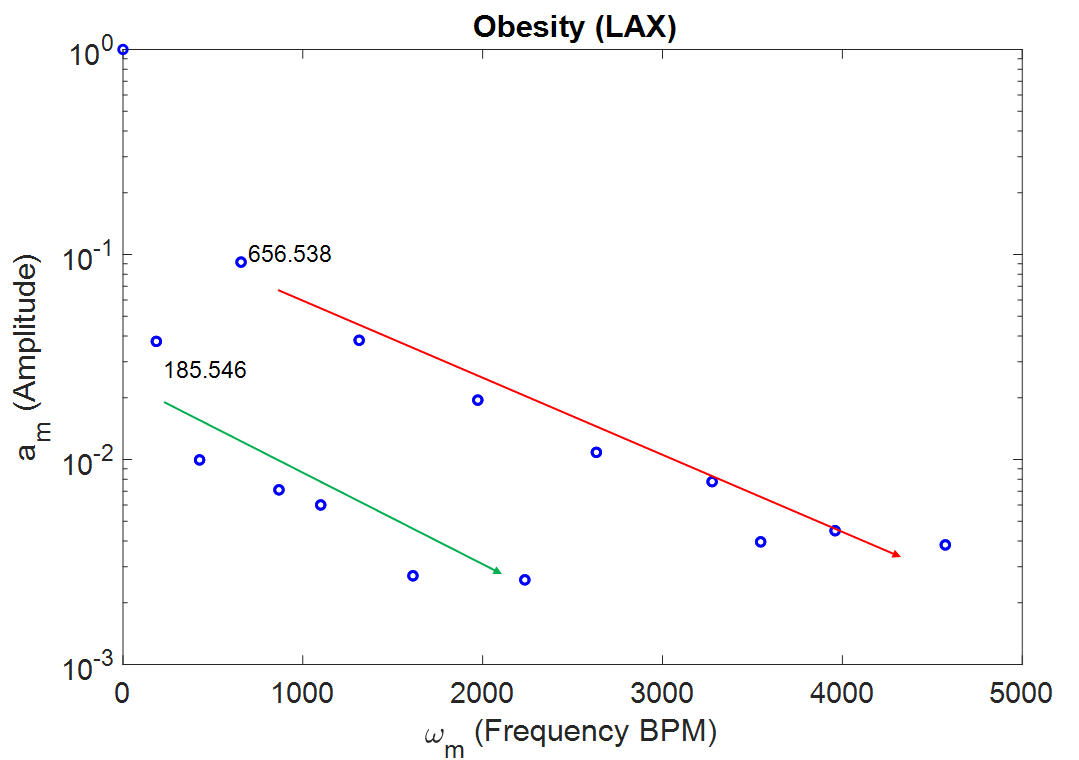}} \qquad
\subfloat[ Frequency of the heart with Obesity (SAX) ]{\includegraphics[width=8.1cm, height=6cm]{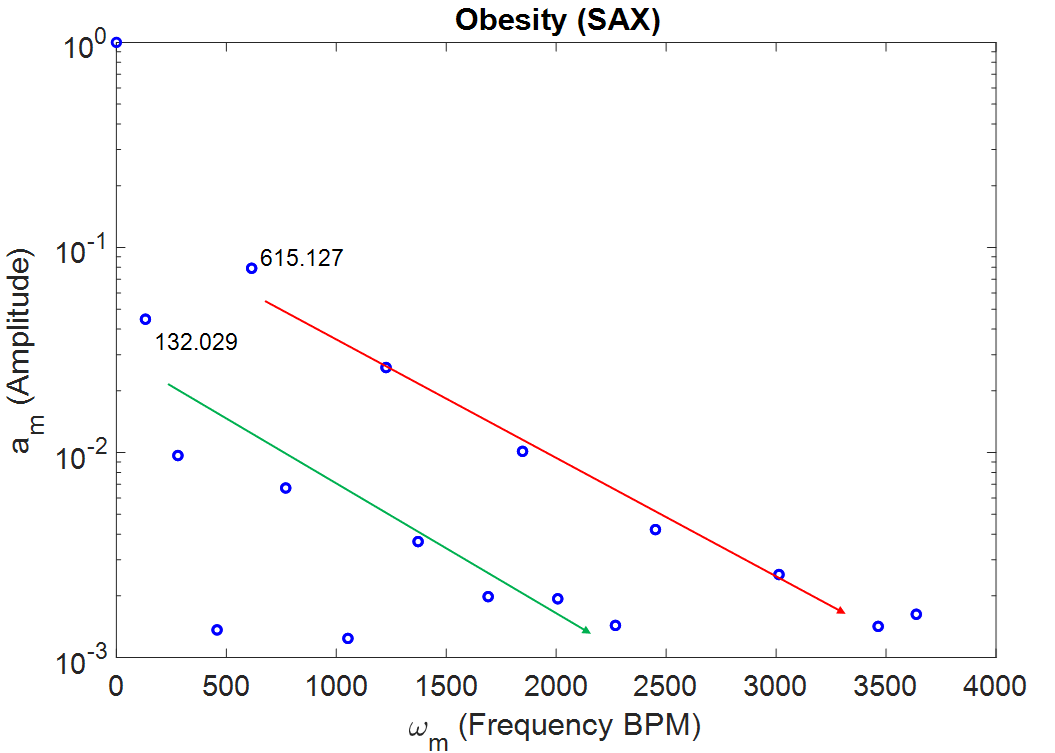}}\\
\subfloat[ Frequency of the heart with TAC Hypertrophy (LAX) ]{\includegraphics[width=8.1cm, height=5.5cm]{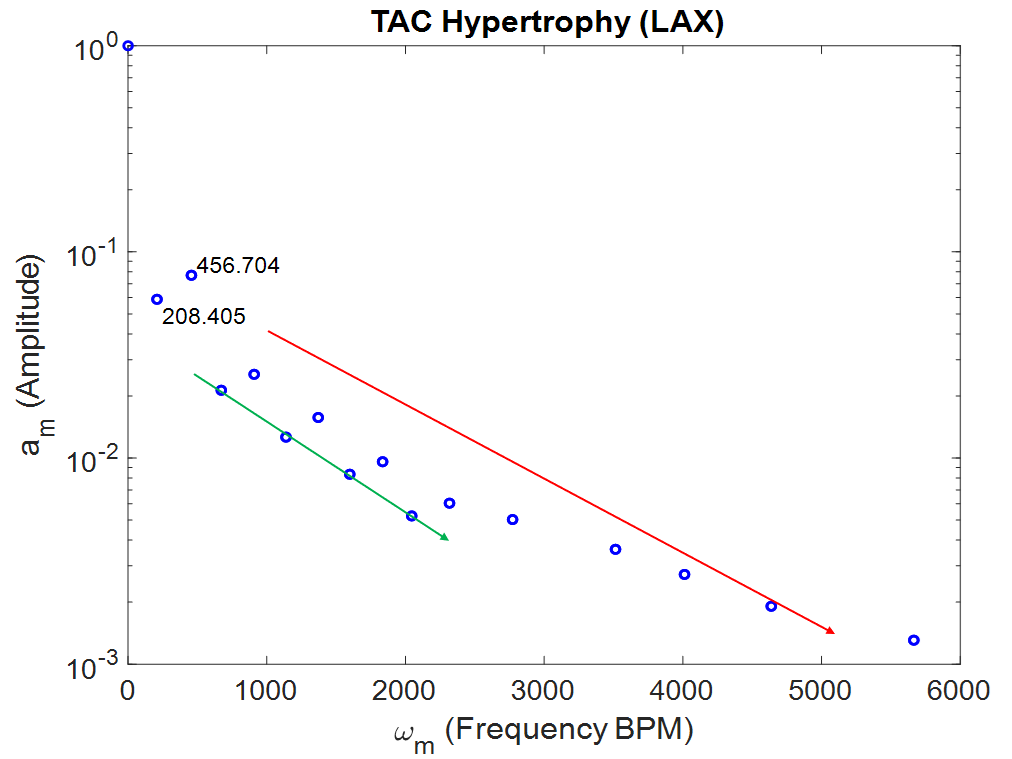}} \qquad
\subfloat[ Frequency of the heart with TAC Hypertrophy (SAX)]{\includegraphics[width=8.1cm, height=5.5cm]{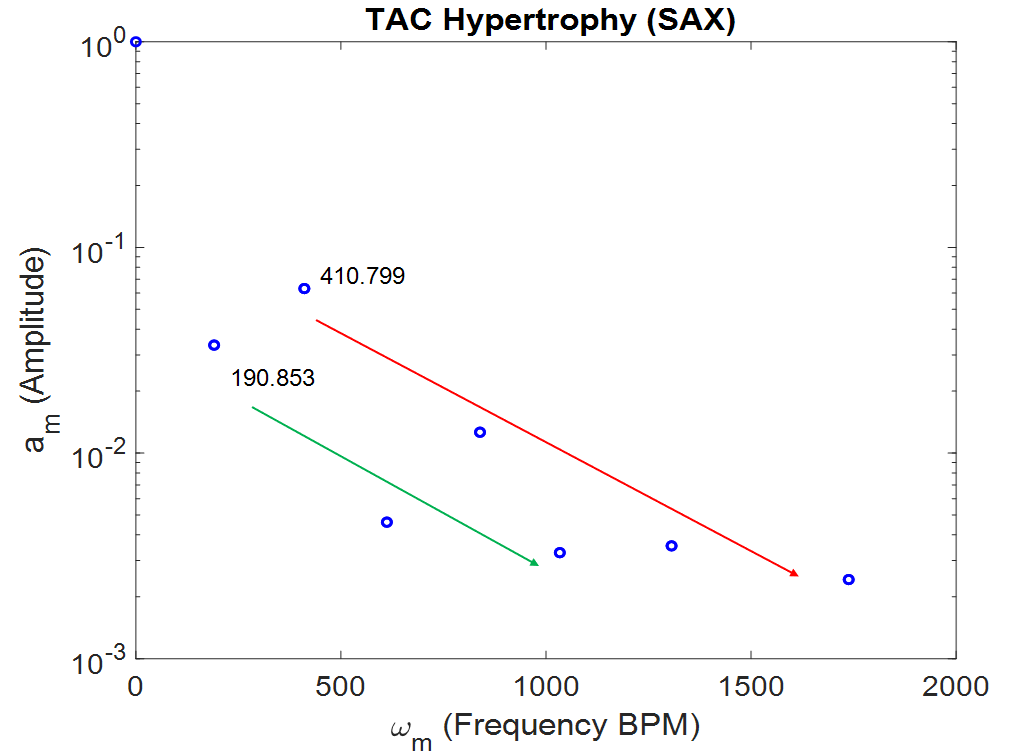}}\\
\subfloat[ Frequency of the heart with SFSR4 Hypertrophy (LAX) ]{\includegraphics[width=8.1cm, height=6cm]{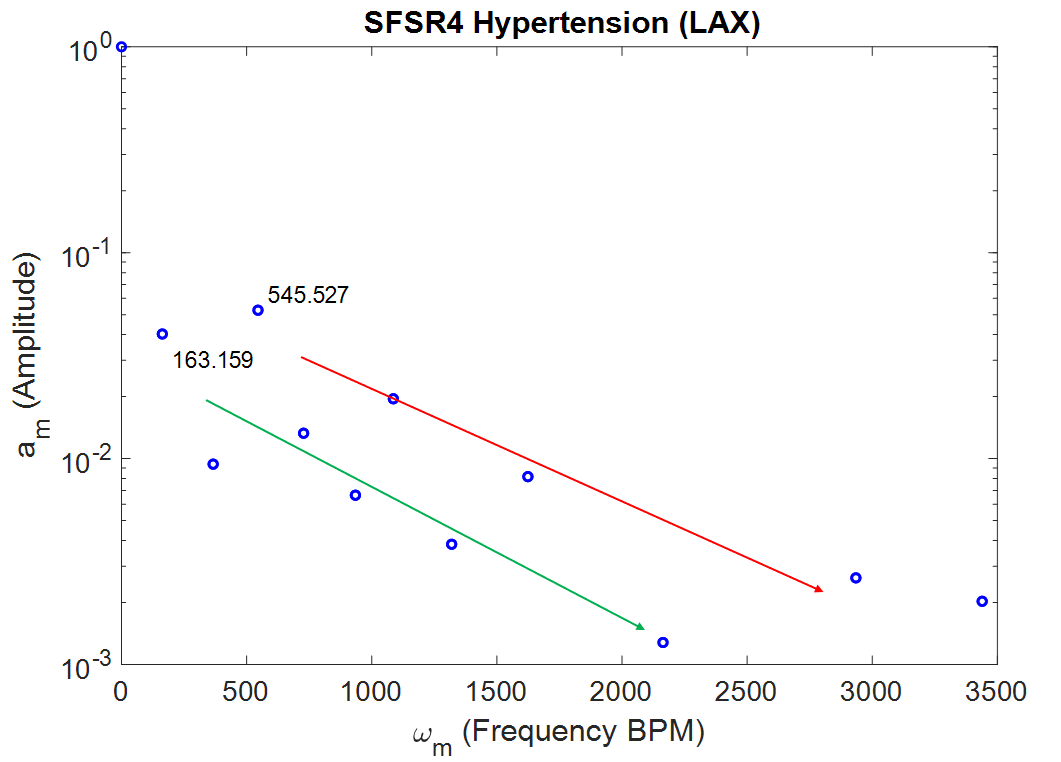}} \qquad
\subfloat[ Frequency of the heart with SFSR4 Hypertrophy (SAX)]{\includegraphics[width=8.1cm, height=6cm]{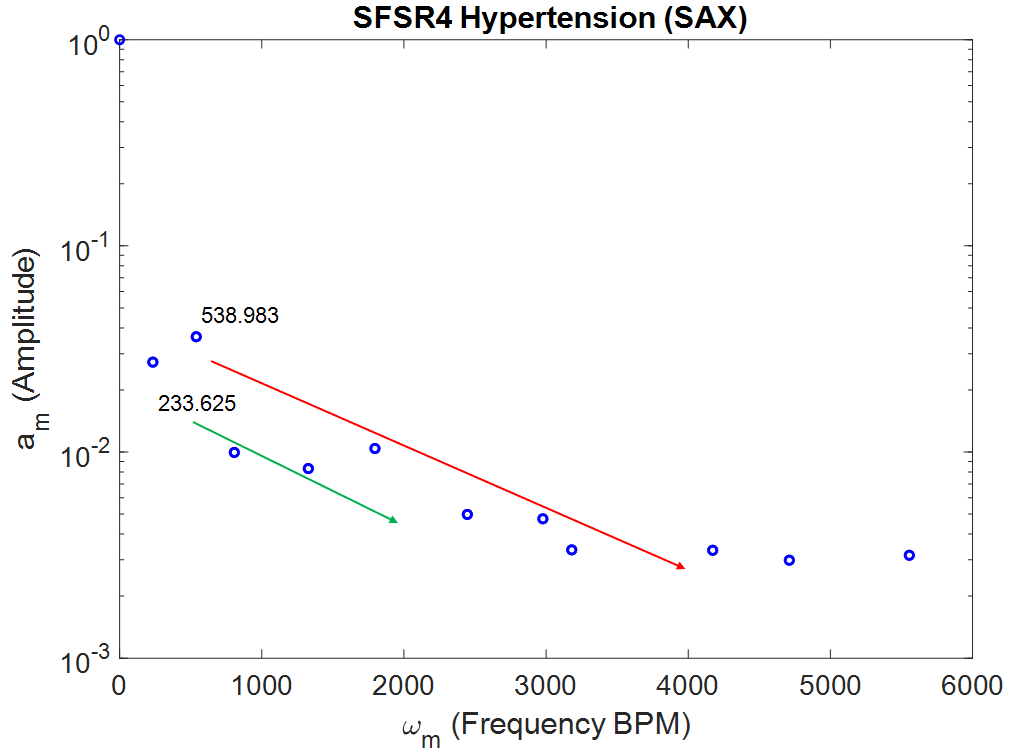}}\\

\centering
	\caption{The frequencies captured from analyzing the data of the hypertrophic hearts. The values of the frequencies from the upper and lower branches can be found in Table \ref{Ob_Freq}, Table \ref{TAC_Freq}, Table \ref{SFSR4_Freq} in Appendix. }
	\label{Hyper_Freq}
\end{figure*}

In the cases of Obesity, TAC Hypertrophy  and SFSR4 Hypertrophy, as seen in Fig. \ref{Hyper_Modes},  all the DMD \textit{modes} show hypertrophy, as they show increased wall thickness compared to the healthy case. This change in thickness is evidenced in the circular shape of the heart and in the left ventricle, which is not as long and thin as in a healthy heart, but instead it is thick and rounded. Regarding the frequencies of the DMD modes presented in Fig. \ref{Hyper_Freq}, they are similar as in the healthy case studied (see all the frequencies in Table \ref{Ob_Freq}, Table \ref{TAC_Freq}, Table \ref{SFSR4_Freq} in Appendix).\\
 \subsubsection{Case 03: Myocardial Infarction model}
\begin{figure*}[h!]
\textbf{Case 03- Myocardial Infarction (Myocardial Infarction model)}\par\medskip
	\centering

\subfloat[ Frequency of the heart with Myocardial Infarction (LAX) ]{\includegraphics[width=8cm, height=6.3cm]{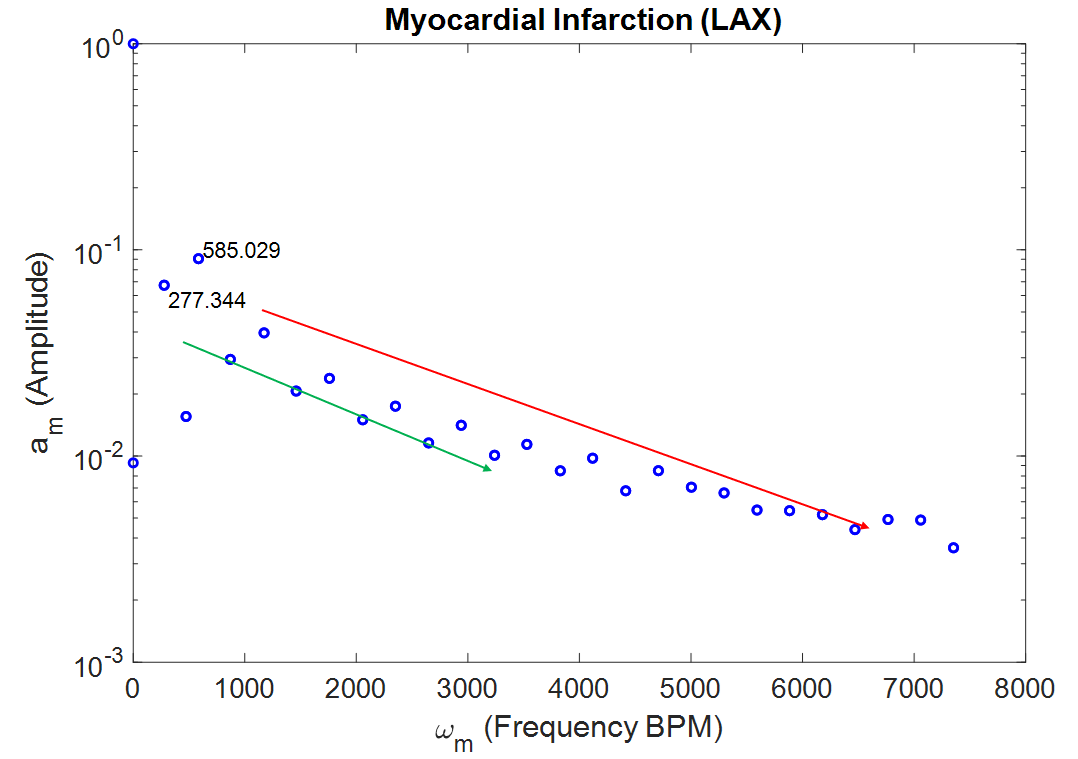}} \qquad
\subfloat[ Frequency of the heart with Myocardial Infarction (SAX) ]{\includegraphics[width=8cm, height=6.3cm]{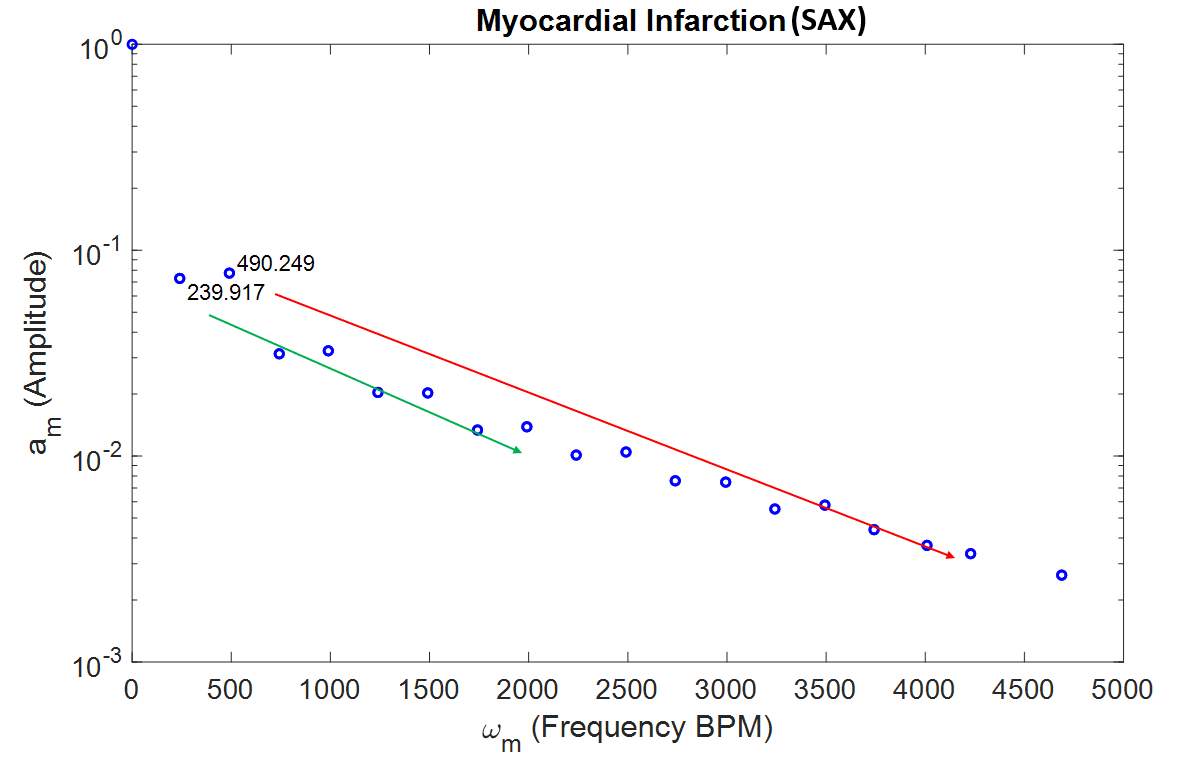}}\\
\subfloat[ Original LAX view ]{\includegraphics[width=4.5cm, height=3.5cm]{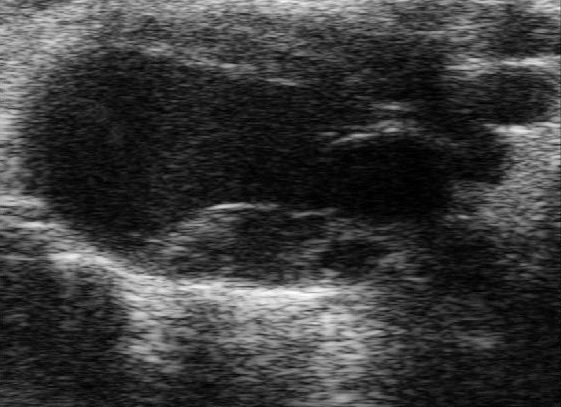}} \hspace{3.3cm}
\subfloat[ Original SAX view ]{\includegraphics[width=4.5cm, height=3.5cm]{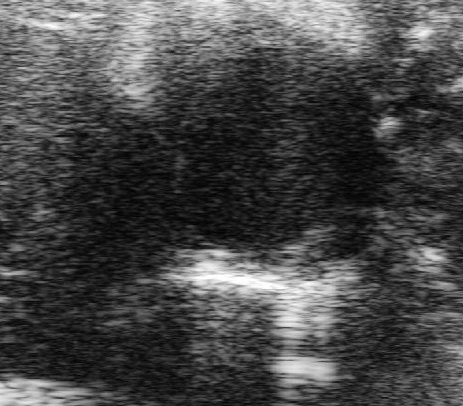}} \\ 
	\subfloat[ Upper branch mode ]{\includegraphics[width=4.4cm, height=3.5cm]{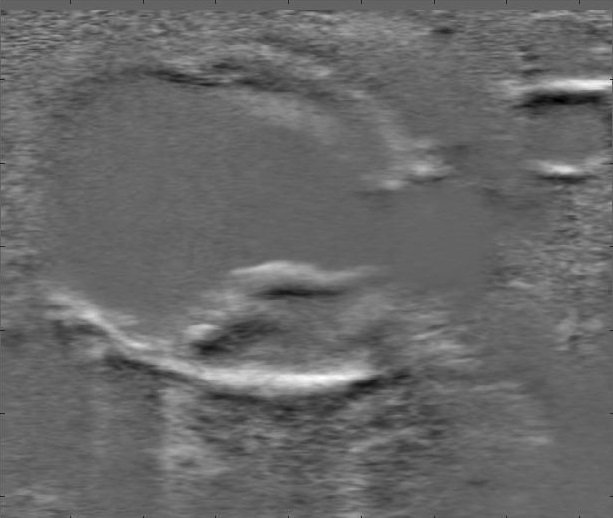}} \hspace{3.3cm}
\subfloat[ Upper branch mode ]{\includegraphics[width=4.5cm, height=3.5cm]{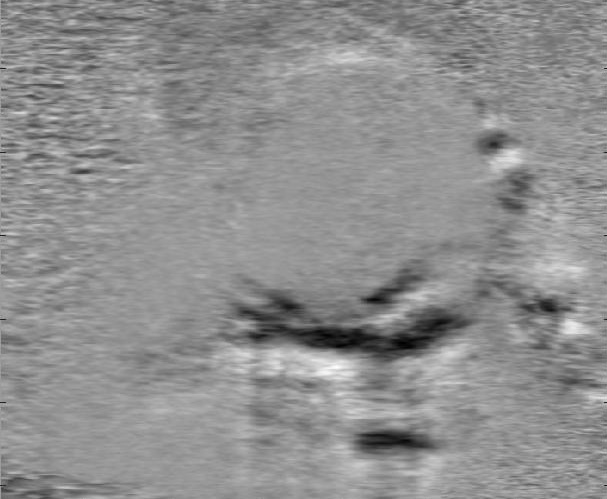}} \\ 
	\subfloat[Lower branch mode ]{\includegraphics[width=4.4cm, height=3.5cm]{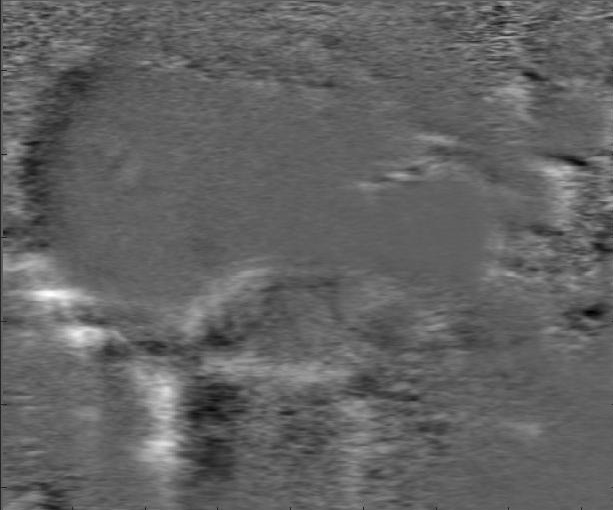}}	\hspace{3.3cm} 
	\subfloat[Lower branch mode ]{\includegraphics[width=4.5cm, height=3.5cm]{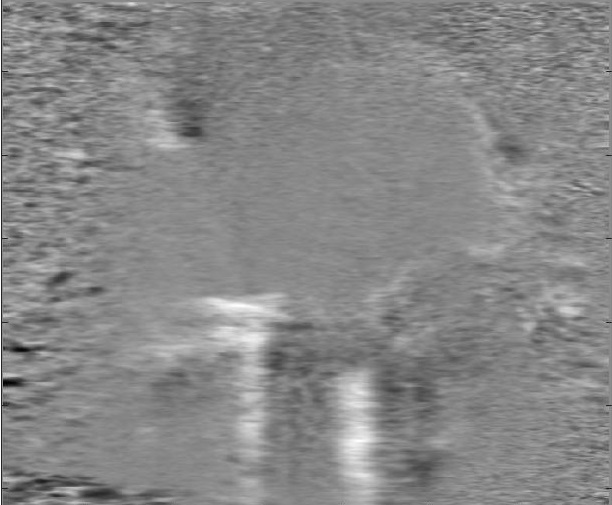}}		
\\
\centering
	\caption{The frequencies captured from analyzing the data of the  hearts suffering from Myocardial Infarction and the dominant modes related to each branch of frequency. The data are normalized with their maximum value. The color-scale correspond to -1 (black), 0 (grey), 1 (white). The values of the frequencies from the upper and lower branches can be found in Table \ref{MI_Freq} in Appendix. }
	\label{MI}
\end{figure*}

Finally, the Myocardial Infarction is presented in Fig. \ref{MI}. From the LAX view, we can clearly see the change in the shape of the heart, as we can notice that the left ventricle does not show the normal shape of a heart, instead it shows a myocardial wall deformation called aneurysm (an outward bulging, likened to a bubble or balloon, caused by a blood vessel obstruction and tissue). Furthermore, the posterior wall is inconspicuous, and it is undetectable because the myocardial thickness is lost due to the infarction. In the SAX view, we can not clearly see the aneurysm, but in both cases the noticeable matter is how the anterior wall is thinner and almost invisible. Thus, we can conclude that the DMD \textit{modes} for both LAX and SAX data, are  clearly capturing all the patterns and features that highly represent this particular cardiac disease. Regarding the frequencies, they are similar as in the healthy case for the LAX data, but they show a slight decrease in the SAX data, which may be explain by a higher impact of the anesthesia in mice with the disease.  (see Table \ref{MI_Freq} in Appendix).  

\vspace{0.5cm}

\FloatBarrier
\subsection{Robustness of the methodology used to identify patterns in medical image}
In addition to the the previous datasets, two extra datasets from each cardiac disease (expect for the SFSR4 hypertrophy) were analyzed. The additional datasets, which are also echocardiography taken from a LAX and SAX view, are composed with a number of snapshots varying between 80 and 135 snapshots. Despite the low number of snapshots, the proposed method  is still able to properly identify the main patterns related to each pathology. In each one of the cases, HODMD captures the two parallel line of frequencies, representing both the heart rate and the respiratory rate. Furthermore, when comparing the frequencies obtained from analyzing data sets belonging to the same cardiac disease, no abnormal differences were noticed, as all the frequencies are varying in a reasonable interval. The resulted DMD modes (not shown for the sake of brevity, but with similar shape as their corresponding previous cases analyzed), as expected, prove the robustness of the HODMD algorithm. The method also captures the aneurysm and the thin appearance of the anterior wall for the myocardial infarction datasets. They show the rounded shape of the heart and thickness of the left ventricle for the hypertension models (Obesity and TAC hypertrophy). As for the DMD modes obtained from the diabetic cardiomyopathy datasets, they are almost identical and they show the same shape and patterns. \balance
\section{Conclusion}\label{Con}
In this work we have investigated, for the first time to the authors knowledge, the use of HODMD for the analysis of medical imaging, in particular, echocardiography images taken from several mice individuals. HODMD, which  was originally developed as a fluid dynamic tool, has been leveraged as a feature extraction technique. The proposed method was initially applied to six different echocardiography datasets, one belonging to a healthy subject and five belonging to subjects afflicted by different cardiac diseases. Each dataset consisted of two video loops, a LAX view and a SAX view, each one composed with a maximum of 300 snapshots, covering at least three cardiac cycles. HODMD was first used to analyze the echocardiography taken from the healthy mouse. To test the limits of applicability of this tool, we have used only 200 snapshots for LAX data and 100 snapshots for SAX data. HODMD has been systematically capable of capturing two parallel lines of frequencies, an upper branch of frequency representing the heart rate and a lower branch representing the respiratory rate. All the frequencies were harmonics, revealing that the solution was periodic, as expected. The DMD modes that belong to the upper branch exhibit the area representing the heart, and the modes that belong to the lower branch exhibit the area representing the lungs. The echocardiography datasets taken from five different mice with cardiac diseases (Diabetic Cardiomyopathy, Obesity, SFSR4, TAC Hypertrophy or myocardial infarction) are analyzed next. As expected, the algorithm still captures the two parallel lines of frequency in every case. The number of modes identified may differ, as well as the frequencies, but this changes are reasonable, since the level of noise in the data, the characteristics of the disease as well as the anesthesia have an impact on the results. The modes obtained represent the dominant features and patterns in each disease, which can be sorted into two categories: (i) the modes obtained from the myocardial infarction echocardiography represent the patterns characterizing the myocardial infarction, which reflect the presence of a myocardial wall deformation (aneurysm) in the left ventricle of the heart, and the invisibility of the anterior wall because of the loss of the myocardial thickness. Meanwhile, (ii) the modes obtained from analyzing obesity, TAC hypertrophy and SFSR4 hypertrophy echocardiography datasets represent patterns that characterize different types of hypertrophy, which are the thickness of the wall and the rounded shape of the heart. Nevertheless, as expected, the DMD modes resulted from analyzing the diabetic cardiomyopathy echocardiography did not show any abnormalities compared to the healthy dataset, and the patterns captured did not display any unusual features that might represent the disease. However, while analyzing this dataset, a clear decrease in the frequencies was noticed (in SAX), which is expected in this disease in particular. Several Additional echocardiography datasets for each disease were later analyzed in order to assess the robustness of our conclusions. The DMD modes obtained were indeed robust and accurate, proving once again the ability of this algorithm to capture spatio-temporal patterns from different, limited, noisy data, in particular to classify cardiac diseases and highlights the usefulness of the HODMD algorithm in other fields besides fluid mechanics. 
\newpage


\clearpage
\appendix


\FloatBarrier
\begin{table}[H]

\begin{center}
\begin{tabular}{|c|c|c|}
\hline
\multirow{2}{2cm}{Frequencies}& \multicolumn{2}{p{2.5cm}|}{\centering Healthy} \\
\cline{2-3} & \multicolumn{1}{c|}{LAX} & \multicolumn{1}{c|}{SAX}  \bigstrut \\ \hline
$\omega_1$ & 0      &     0  \bigstrut \\ \hline
$\omega_2$ & 208.6 & 203.8   \bigstrut \\ \hline
$\omega_3$ & 425.4 & 419.0 \bigstrut  \\ \hline
$\omega_4$ & 633.4 & 644.1  \bigstrut  \\ \hline
$\omega_5$ & 859.2 &  856.2 \bigstrut \\ \hline
$\omega_6$ & 1074.2 & 1063.9  \bigstrut \\ \hline
$\omega_7$ & 1273.2 &  1289.2 \bigstrut  \\ \hline
$\omega_8$ & 1502.4 &  1505.0 \bigstrut \\ \hline
$\omega_9$ & 1720.6 &  1716.4 \bigstrut \\ \hline
$\omega_{10}$ & 1916.9 &  1934.4 \bigstrut  \\ \hline
$\omega_{11}$ & 2144.6 &  2203.4  \bigstrut \\ \hline
$\omega_{12}$ & 2365.7 & 2582.6 \bigstrut \\ \hline
$\omega_{13}$ & 2560.7 &  2735.8 \bigstrut  \\ \hline
$\omega_{14}$ & 2786.9 & 3230.1 \bigstrut \\ \hline
$\omega_{15}$ & 3009.0 & 3889.9  \bigstrut \\ \hline
$\omega_{16}$ & 3203.3 & 4526.7  \bigstrut  \\ \hline
$\omega_{17}$ & 3428.3 &  5186.9 \bigstrut \\ \hline
$\omega_{18}$ & 3652.0 & 5725.5  \bigstrut \\ \hline
$\omega_{19}$ &    & 7499.9  \bigstrut  \\
\hline
\end{tabular}
\end{center}
\caption{Frequency of the healthy data set.}
\label{H_Freq}
\end{table}
%
%
%
%
\FloatBarrier
\begin{table}[H]
\begin{center}
\begin{tabular}{|c|c|c|}
\hline
\multirow{2}{2cm}{Frequencies}& \multicolumn{2}{p{2.5cm}|}{\centering Diabetic Cardiomyopathy} \\
\cline{2-3} & \multicolumn{1}{c|}{LAX} & \multicolumn{1}{c|}{SAX}  \bigstrut \\ \hline
$\omega_1$ & 0      &     0  \bigstrut \\ \hline
$\omega_2$ & 213.4 & 196.7   \bigstrut \\ \hline
$\omega_3$ & 493.2 & 353.8 \bigstrut  \\ \hline
$\omega_4$ & 747.4 & 494.7  \bigstrut  \\ \hline
$\omega_5$ & 984.3 &  703.2 \bigstrut \\ \hline
$\omega_6$ & 1240.6 & 819.7  \bigstrut \\ \hline
$\omega_7$ & 1489.9 &  997.7 \bigstrut  \\ \hline
$\omega_8$ & 1707.0 &  1284.6 \bigstrut \\ \hline
$\omega_9$ & 1999.6 & 1500.8 \bigstrut \\ \hline
$\omega_{10}$ & 2227.1 & 1902.9  \bigstrut  \\ \hline
$\omega_{11}$ & 2475.5 & 1995.4  \bigstrut \\ \hline
$\omega_{12}$ & 3003.2 & 2505.0  \bigstrut \\ \hline
$\omega_{13}$ & 3506.5 & 3003.3  \bigstrut  \\ \hline
$\omega_{14}$ & 3750 & 3505.1  \bigstrut \\ \hline
$\omega_{15}$ &        & 3984.0  \bigstrut \\ \hline
$\omega_{16}$ &        & 4493.7   \bigstrut  \\ 

\hline
\end{tabular}
\end{center}
\caption{Frequency of the Diabetic Cardiomyopathy data set.}
\label{DC_Freq}
\end{table}
%
\FloatBarrier
\begin{table}[H]
\begin{center}
\begin{tabular}{|c|c|c|}
\hline

\multirow{2}{2cm}{Frequencies}& \multicolumn{2}{p{2.5cm}|}{\centering Obesity} \\
\cline{2-3} & \multicolumn{1}{c|}{LAX} & \multicolumn{1}{c|}{SAX}  \bigstrut \\ \hline
$\omega_1$ & 0      &     0  \bigstrut \\ \hline
$\omega_2$ & 185.5 & 132.0   \bigstrut \\ \hline
$\omega_3$ & 425.9 & 279.6 \bigstrut  \\ \hline
$\omega_4$ & 656.5 & 457.7  \bigstrut  \\ \hline
$\omega_5$ & 867.5 &  615.1 \bigstrut \\ \hline
$\omega_6$ & 1099.5 & 770.0  \bigstrut \\ \hline
$\omega_7$ & 1313.2 &  1052.8 \bigstrut  \\ \hline
$\omega_8$ & 1612.2 &  1226.1 \bigstrut \\ \hline
$\omega_9$ & 1972.8 &  1371.9 \bigstrut \\ \hline
$\omega_{10}$ & 2234.4 &  1690.5 \bigstrut  \\ \hline
$\omega_{11}$ & 2632.0 &  1846.8 \bigstrut \\ \hline 
$\omega_{12}$ & 3275.6 & 2006.9 \bigstrut \\ \hline
$\omega_{13}$ & 3545.7 &  2269.2 \bigstrut  \\  \hline
$\omega_{14}$ & 3959.5 & 2451.2 \bigstrut \\ \hline
$\omega_{15}$ & 4572.2 & 3012.9  \bigstrut \\ \hline
$\omega_{16}$ &     & 3464.0  \bigstrut  \\ \hline
$\omega_{17}$ &     &  3637.3 \bigstrut \\ 

\hline
\end{tabular}
\end{center}
\caption{Frequency of the Obesity data set.}
\label{Ob_Freq}
\end{table}

\FloatBarrier
\begin{table}[h!]
\begin{center}
\begin{tabular}{|c|c|c|}
\hline

\multirow{2}{2cm}{Frequencies}& \multicolumn{2}{p{2.5cm}|}{\centering TAC Hypertrophy} \\
\cline{2-3} & \multicolumn{1}{c|}{LAX} & \multicolumn{1}{c|}{SAX}  \bigstrut \\ \hline
$\omega_1$ & 0      &     0  \bigstrut \\ \hline
$\omega_2$ & 208.4 & 190.8   \bigstrut \\ \hline
$\omega_3$ & 456.7 & 410.7 \bigstrut  \\ \hline
$\omega_4$ & 671.9 & 612.0  \bigstrut  \\ \hline
$\omega_5$ & 908.9 &  839.0 \bigstrut \\ \hline
$\omega_6$ & 1138.5 & 1033.7  \bigstrut \\ \hline
$\omega_7$ & 1371.0 &  1306.1 \bigstrut  \\ \hline
$\omega_8$ & 1598.5 &  1737.9 \bigstrut \\ \hline
$\omega_9$ & 1835.3 &   \bigstrut \\ \hline
$\omega_{10}$ & 2045.0 &  \bigstrut  \\ \hline
$\omega_{11}$ & 2317.5 &   \bigstrut \\ \hline
$\omega_{12}$ & 2772.3 &  \bigstrut \\ \hline
$\omega_{13}$ & 3514.4 &   \bigstrut  \\ \hline
$\omega_{14}$ & 4011.9 &  \bigstrut \\ \hline 
$\omega_{15}$ & 4637.4 &  \bigstrut \\ \hline
$\omega_{16}$ &  5665.3   &  \bigstrut  \\ 

\hline
\end{tabular}
\end{center}
\caption{Frequency of the TAC Hypertrophy data set.}
\label{TAC_Freq}
\end{table}

\FloatBarrier
\begin{table}[h!]
\begin{center}
\begin{tabular}{|c|c|c|}
\hline
\multirow{2}{2cm}{Frequencies}& \multicolumn{2}{p{2.5cm}|}{\centering SFSR4 Hypertrophy} \\
\cline{2-3} & \multicolumn{1}{c|}{LAX} & \multicolumn{1}{c|}{SAX}  \bigstrut \\ \hline
$\omega_1$ & 0      &     0  \bigstrut \\ \hline
$\omega_2$ & 163.1 & 233.6   \bigstrut \\ \hline
$\omega_3$ & 366.3 & 538.9 \bigstrut  \\ \hline
$\omega_4$ & 545.5 & 806.5  \bigstrut  \\ \hline
$\omega_5$ & 727.7 &  934.8 \bigstrut \\ \hline
$\omega_6$ & 934.8 & 1327.3  \bigstrut \\ \hline
$\omega_7$ & 1086.3 &  2445.8 \bigstrut  \\ \hline
$\omega_8$ & 1319.4 &  2977.2 \bigstrut \\ \hline
$\omega_9$ & 1624.0 & 3180.4   \bigstrut \\ \hline
$\omega_{10}$ & 2163.9 & 4171.9 \bigstrut  \\ \hline
$\omega_{11}$ &2934.6 &  4710.8 \bigstrut \\ \hline
$\omega_{12}$ & 3439.2 &  5554.8\bigstrut \\ 

\hline
\end{tabular}
\end{center}
\caption{Frequency of the SFSR4 Hypertrophy data set.}
\label{SFSR4_Freq}
\end{table}

\FloatBarrier
\begin{table}[h!]
\begin{center}
\begin{tabular}{|c|c|c|}
\hline
\multirow{2}{2cm}{Frequencies}& \multicolumn{2}{p{2.5cm}|}{\centering Myocardial Infarction} \\
\cline{2-3} & \multicolumn{1}{c|}{LAX} & \multicolumn{1}{c|}{SAX}  \bigstrut \\ \hline
$\omega_1$ & 0      &     0  \bigstrut \\ \hline
$\omega_2$ & 277.3 & 239.9   \bigstrut \\ \hline
$\omega_3$ & 473.6 & 490.2 \bigstrut  \\ \hline
$\omega_4$ & 585.0 & 741.9 \bigstrut  \\ \hline
$\omega_5$ & 870.4 &  989.2 \bigstrut \\ \hline
$\omega_6$ & 1173.0 & 1239.5  \bigstrut \\ \hline
$\omega_7$ & 1460.8 & 1490.6 \bigstrut  \\ \hline
$\omega_8$ & 1759.1 &  1741.6 \bigstrut \\ \hline
$\omega_9$ & 2057.4 &  1990.7 \bigstrut \\ \hline
$\omega_{10}$ & 2350.3 &  2239.5 \bigstrut  \\ \hline
$\omega_{11}$ & 2648.2 &  2490.6  \bigstrut \\ \hline
$\omega_{12}$ & 2940.3 & 2739.0 \bigstrut \\ \hline
$\omega_{13}$ & 3238.9 &  2993.5 \bigstrut  \\ \hline
$\omega_{14}$ & 3528.7 & 3241.4 \bigstrut \\ \hline
$\omega_{15}$ & 3528.7 & 3241.4  \bigstrut \\ \hline
$\omega_{16}$ & 3828.4 & 3493.7  \bigstrut  \\ \hline
$\omega_{17}$ & 4117.9 &  3741.8 \bigstrut \\ \hline
$\omega_{18}$ & 4414.8 & 4008.5  \bigstrut \\ \hline
$\omega_{19}$ & 4707.8   & 4228.9  \bigstrut  \\
\hline
$\omega_{20}$ & 5003.1   & 4688.1  \bigstrut  \\
\hline
$\omega_{21}$ &  5296.6  &   \bigstrut  \\
\hline
$\omega_{22}$ &  5591.7  &   \bigstrut  \\
\hline
$\omega_{23}$ &  5883.5  &   \bigstrut  \\
\hline
$\omega_{24}$ &  6177.9  &   \bigstrut  \\
\hline
$\omega_{25}$ &  6469.7  &   \bigstrut  \\
\hline
$\omega_{26}$ &  6765.7  &   \bigstrut  \\
\hline
$\omega_{27}$ &  7058.6  &   \bigstrut  \\
\hline
$\omega_{28}$ &  7352.9  &   \bigstrut  \\
\hline

\end{tabular}
\end{center}
\caption{Frequency of the Myocardial Infarction data set.}
\label{MI_Freq}
\end{table}

\newpage

 \pagestyle{empty}

\newcolumntype{C}{>{\centering\arraybackslash}p{2.7em}}
\begin{landscape}


\newcolumntype{g}{>{\columncolor{Gray}}c}
\FloatBarrier
\begin{table*}[h!]

\begin{center}
\resizebox{20cm}{!}{
\hspace*{-2.9in}\begin{tabular}{|g|g|c|c|c|c|c|c|c|c|c|c|c|c|c|c|c|c|c|c|c|c|}
\midrule
\rowcolor{Gray}
\multicolumn{2}{|c|}{Frequencies}& $\omega_1$  & $\omega_2$ & $\omega_3$ & $\omega_4$ & $\omega_5$ & $\omega_6$  & $\omega_7$ & $\omega_8$ & $\omega_9$ & $\omega_{10}$ & $\omega_{11}$ & $\omega_{12}$ & $\omega_{13}$ &  $\omega_{14}$ & $\omega_{15}$ & $\omega_{16}$ & $\omega_{17}$ & $\omega_{18}$ &  $\omega_{19}$ & $\omega_{20}$ \bigstrut \\
 \midrule[1pt]
\multirow{2}{*}{Healthy} &  LAX  & 0  & 208.6 & 425.4 & 633.4 & 859.2 & 1074.2 & 1273.2 & 1502.4 & 1720.6 & 1916.9 & 2144.6 & 2365.7 & 2560.7 & 2786.9 &  3009.0 &3203.3 & 3428.3 & 3652.0 &  &   \bigstrut \\
\cline{2-22}
 & SAX & 0 & 203.8  & 419.0 & 644.1 & 856.2 & 1063.9 & 1289.2  & 1505.0 & 1716.4 & 1934.4 & 2203.4 & 2582.6 & 2735.8 & 3230.1 &  3889.9 & 4526.7 & 5186.9 & 5725.5 & 7499.9 &       \bigstrut \\
 \midrule[1pt]
\multirow{2}{*}{Diabetic } &  LAX  & 0  & 213.4 & 493.2 & 747.4 & 984.3 & 1240.6 & 1489.9 & 1707.0 & 1999.6 & 2227.1 & 2475.5 & 3003.2 & 3506.5 &  3750 & &  &  &  &   &    \bigstrut \\
\cline{2-22}
 Cardiomyopathy & SAX & 0  & 196.7 & 353.8 & 494.7 & 703.2 & 819.7 & 997.7 &1284.6  & 1500.8  & 1500.8 & 1902.9 & 1995.4 & 2505.0 & 3003.3 &  3505.1 & 3984.0 & 4493.7 & &  &      \bigstrut \\
 \midrule[1pt]
\multirow{2}{*}{Obesity} &  LAX  & 0  & 185.5 & 425.9 & 656.5 & 867.5 & 1099.5 & 1313.2 & 1612.2 & 1972.8 & 2234.4 & 2632.0 & 3275.6 & 3545.7 &  3959.5 & 4572.2 &  &  &  &   &    \bigstrut \\
\cline{2-22} & SAX & 0  & 132.0 & 279.6 & 457.7 & 615.1 & 770.0  & 1052.8 & 1226.1 & 1371.9 & 1690.5 & 1846.8 & 2006.9 & 2269.2 &  2451.2 & 3012.9 & 3464.0 & 3637.3 &  &  &  \bigstrut \\ 
\midrule[1pt]
\multirow{2}{*}{TAC} &  LAX  & 0  & 208.4 & 456.7 & 671.9 & 908.9 & 1138.5  & 1371.0 & 1598.5 & 1835.3 & 2045.0 & 2317.5 & 2772.3 & 3514.4 &  4011.9 & 4637.4 & 5665.3 &  &  &   &    \bigstrut \\
\cline{2-22}
Hypertrophy & SAX & 0  & 190.8 & 410.7 &612.0 & 839.0 & 1033.7  & 1306.1 & 1737.9 &  &  &  &  &  &   &  &  &  &  &   &   \bigstrut \\
 \midrule[1pt]
\multirow{2}{*}{SFSR4} &  LAX  & 0  & 163.1 & 366.3 & 545.5 & 727.7 & 934.8  & 1086.3 & 1319.4 & 1624.0 & 2163.9 & 2934.6 & 3439.2 &  &  &  &  &  &  &   &  \bigstrut \\
\cline{2-22}
Hypertrophy & SAX & 0  & 233.6 & 538.9 & 806.5 &  1327.3 & 2445.8 & 2977.2 & 3180.4 & 4171.9 & 4710.8 & 5554.8 &  &  &  &  &  &  &  &  & \bigstrut \\
 \midrule[1pt]
\multirow{2}{*}{Myocardial} &  LAX  & 0  & 277.3 & 473.6 & 585.0 & 870.4 & 1173.0  & 1460.8 & 1759.1 & 2057.4 & 2350.3 & 2648.2 & 2940.3 & 3238.9 &  3528.7 & 3828.4 & 4117.9 & 4414.8 & 4707.8 &  5003.1 & 5296.6 \bigstrut \\
\cline{2-22}
Infarction & SAX & 0  & 239.9 & 490.2 & 741.9 & 989.2 & 1239.5  & 1490.6 & 1741.6 & 1990.7 & 2239.5 & 2490.6 & 2739.0 & 2993.5 &  3241.4 & 3493.7 & 3741.8 & 4008.5 & 4228.9 &  4688.1 &   \bigstrut \\ \hline

\end{tabular}
} 

\caption{Frequency of the DMD modes.}
\label{TAB_Freq}
\end{center}
\end{table*} 

\end{landscape}

\end{document}